\documentclass[twocolumn,10pt,showpacs,showkeys,preprintnumbers,amsmath,amssymb,floatfix,aps,prc]{revtex4-1}
\usepackage[latin9]{inputenc}
\usepackage{color}
\usepackage{graphicx}

\usepackage[colorlinks,urlcolor=blue]{hyperref}
\usepackage{breakurl}

\makeatletter

\@ifundefined{textcolor}{}
{%
 \definecolor{BLACK}{gray}{0}
 \definecolor{WHITE}{gray}{1}
 \definecolor{RED}{rgb}{1,0,0}
 \definecolor{GREEN}{rgb}{0,1,0}
 \definecolor{BLUE}{rgb}{0,0,1}
 \definecolor{CYAN}{cmyk}{1,0,0,0}
 \definecolor{MAGENTA}{cmyk}{0,1,0,0}
 \definecolor{YELLOW}{cmyk}{0,0,1,0}
}

\newcounter{univ_counter}
\addtocounter{univ_counter} {1} 
\edef\ANL{$^{\arabic{univ_counter}}$ }

\addtocounter{univ_counter} {1} 
\edef\ASU{$^{\arabic{univ_counter}}$ }

\addtocounter{univ_counter} {1} 
\edef\BRESCIA{$^{\arabic{univ_counter}}$ }

\addtocounter{univ_counter} {1} 
\edef\UCR{$^{\arabic{univ_counter}}$ }

\addtocounter{univ_counter} {1} 
\edef\CSUDH{$^{\arabic{univ_counter}}$ }

\addtocounter{univ_counter} {1} 
\edef\CANISIUS{$^{\arabic{univ_counter}}$ }

\addtocounter{univ_counter} {1} 
\edef\CNU{$^{\arabic{univ_counter}}$ }

\addtocounter{univ_counter} {1} 
\edef\UCONN{$^{\arabic{univ_counter}}$ }

\addtocounter{univ_counter} {1} 
\edef\DUKE{$^{\arabic{univ_counter}}$ }

\addtocounter{univ_counter} {1} 
\edef\DUQUESNE{$^{\arabic{univ_counter}}$ }

\addtocounter{univ_counter} {1} 
\edef\FU{$^{\arabic{univ_counter}}$ }

\addtocounter{univ_counter} {1}
\edef\FERRARAU{$^{\arabic{univ_counter}}$ }

\addtocounter{univ_counter} {1} 
\edef\FIU{$^{\arabic{univ_counter}}$ }

\addtocounter{univ_counter} {1} 
\edef\GWUI{$^{\arabic{univ_counter}}$ }

\addtocounter{univ_counter} {1} 
\edef\JLUGIESSEN{$^{\arabic{univ_counter}}$ }

\addtocounter{univ_counter} {1} 
\edef\GSIFFN{$^{\arabic{univ_counter}}$ }

\addtocounter{univ_counter} {1} 
\edef\GLASGOW{$^{\arabic{univ_counter}}$ }

\addtocounter{univ_counter} {1} 
\edef\INFNCAT{$^{\arabic{univ_counter}}$ }

\addtocounter{univ_counter} {1} 
\edef\INFNFE{$^{\arabic{univ_counter}}$ }

\addtocounter{univ_counter} {1} 
\edef\INFNFR{$^{\arabic{univ_counter}}$ }

\addtocounter{univ_counter} {1} 
\edef\INFNGE{$^{\arabic{univ_counter}}$ }

\addtocounter{univ_counter} {1} 
\edef\INFNPAV{$^{\arabic{univ_counter}}$ }

\addtocounter{univ_counter} {1} 
\edef\INFNRO{$^{\arabic{univ_counter}}$ }

\addtocounter{univ_counter} {1} 
\edef\INFNTUR{$^{\arabic{univ_counter}}$ }

\addtocounter{univ_counter} {1} 
\edef\JMU{$^{\arabic{univ_counter}}$ }

\addtocounter{univ_counter} {1} 
\edef\KNU{$^{\arabic{univ_counter}}$ }

\addtocounter{univ_counter} {1} 
\edef\LAMAR{$^{\arabic{univ_counter}}$ }

\addtocounter{univ_counter} {1} 
\edef\MIT{$^{\arabic{univ_counter}}$ }

\addtocounter{univ_counter} {1} 
\edef\MISS{$^{\arabic{univ_counter}}$ }

\addtocounter{univ_counter} {1} 
\edef\UNH{$^{\arabic{univ_counter}}$ }

\addtocounter{univ_counter} {1} 
\edef\NMSU{$^{\arabic{univ_counter}}$ }

\addtocounter{univ_counter} {1} 
\edef\NSU{$^{\arabic{univ_counter}}$ }

\addtocounter{univ_counter} {1} 
\edef\OHIOU{$^{\arabic{univ_counter}}$ }

\addtocounter{univ_counter} {1} 
\edef\ODU{$^{\arabic{univ_counter}}$ }

\addtocounter{univ_counter} {1} 
\edef\ORSAY{$^{\arabic{univ_counter}}$ }

\addtocounter{univ_counter} {1} 
\edef\ROMAII{$^{\arabic{univ_counter}}$ }

\addtocounter{univ_counter} {1} 
\edef\SACLAY{$^{\arabic{univ_counter}}$ }

\addtocounter{univ_counter} {1} 
\edef\MSU{$^{\arabic{univ_counter}}$ }

\addtocounter{univ_counter} {1} 
\edef\SCAROLINA{$^{\arabic{univ_counter}}$ }

\addtocounter{univ_counter} {1} 
\edef\TEMPLE{$^{\arabic{univ_counter}}$ }

\addtocounter{univ_counter} {1} 
\edef\UTFSM{$^{\arabic{univ_counter}}$ }

\addtocounter{univ_counter} {1}
\edef\JLAB{$^{\arabic{univ_counter}}$ }

\addtocounter{univ_counter} {1}
\edef\ULS{$^{\arabic{univ_counter}}$ }

\addtocounter{univ_counter} {1} 
\edef\YEREVAN{$^{\arabic{univ_counter}}$ }

\addtocounter{univ_counter} {1} 
\edef\YORK{$^{\arabic{univ_counter}}$ }

\begin{document}

\preprint{Phys. Rev. C}

% Report #: JLAB-PHY-25-4325

\title{Recoil Polarization in $K^+Y$ Electroproduction in the Nucleon Resonance Region with CLAS12}

\author{D.S.~Carman$^\dag$,\JLAB\
A.~D'Angelo,\INFNRO$\!\!^,$\ROMAII\
L.~Lanza,\INFNRO$\!\!^,$\ROMAII\
V.I.~Mokeev,\JLAB\
P.~Achenbach,\JLAB\
J.S.~Alvarado,\ORSAY\
M.J.~Amaryan,\ODU\
H.~Atac,\TEMPLE\ 
H.~Avakian,\JLAB\
N.A.~Baltzell,\JLAB\
L.~Barion,\INFNFE
% ...
\\
(CLAS Collaboration)}

\affiliation{\ANL Argonne National Laboratory, Argonne, Illinois 60439}
\affiliation{\ASU Arizona State University, Tempe, Arizon 85287}
\affiliation{\BRESCIA Universit\`{a} degli Studi di Brescia, 25123 Brescia, Italy}
\affiliation{\UCR University of California Riverside, 900 University Avenue, Riverside, California 92521}
\affiliation{\CSUDH California State University, Dominguez Hills, Carson, California 90747}
\affiliation{\CANISIUS Canisius College, Buffalo, New York 14208}
\affiliation{\CNU Christopher Newport University, Newport News, VA 23606}
\affiliation{\UCONN University of Connecticut, Storrs, Connecticut 06269}
\affiliation{\DUKE Duke University, Durham, North Carolina 27708}
\affiliation{\DUQUESNE Duquesne University, Pittsburgh, PA 15282}  
\affiliation{\FU Fairfield University, Fairfield, Connecticut 06824}
\affiliation{\FERRARAU Universit\`{a} di Ferrara, 44121 Ferrara, Italy}
\affiliation{\FIU Florida International University, Miami, Florida 33199}
\affiliation{\GWUI The George Washington University, Washington, D.C. 20052}
\affiliation{\JLUGIESSEN II Physikalisches Institut der Universitaet Giessen, 35392 Giessen, Germany}
\affiliation{\GSIFFN GSI Helmholtzzentrum fur Schwerionenforschung GmbH, D-64291 Darmstadt, Germany}
\affiliation{\GLASGOW University of Glasgow, Glasgow G12 8QQ, United Kingdom}
\affiliation{\INFNCAT INFN, Sezione di Catania, 95123 Catania, Italy}
\affiliation{\INFNFE INFN, Sezione di Ferrara, 44100 Ferrara, Italy}
\affiliation{\INFNFR INFN, Laboratori Nazionali di Frascati, 00044 Frascati, Italy}
\affiliation{\INFNGE INFN, Sezione di Genova, 16146 Genova, Italy}
\affiliation{\INFNPAV INFN,  Sezione di Pavia, 27100 Pavia, Italy}
\affiliation{\INFNRO INFN, Sezione di Roma Tor Vergata, 00133 Rome, Italy}
\affiliation{\INFNTUR INFN, Sezione di Torino, 10125 Torino, Italy}
\affiliation{\JMU James Madison University, Harrisonburg, Virginia 22807}
\affiliation{\KNU Kyungpook National University, Daegu 702-701, Republic of Korea}
\affiliation{\LAMAR Lamar University, 4400 MLK Blvd, Beaumont, Texas 77710}
\affiliation{\MIT Massachusetts Institute of Technology, Cambridge, Massachusetts 02139}
\affiliation{\MISS Mississippi State University, Mississippi State, Mississippi 39762}
\affiliation{\UNH University of New Hampshire, Durham, New Hampshire 03824}
\affiliation{\NMSU New Mexico State University, Las Cruces, New Mexico 88003}
\affiliation{\NSU Norfolk State University, Norfolk, Virginia 23504}
\affiliation{\OHIOU Ohio University, Athens, Ohio 45701}
\affiliation{\ODU Old Dominion University, Norfolk, Virginia 23529}
\affiliation{\ORSAY Universit\'{e} Paris-Saclay, CNRS/IN2P3, IJCLab, 91405 Orsay, France}
\affiliation{\ROMAII Universit\`{a} di Roma Tor Vergata, 00133 Rome, Italy}
\affiliation{\SACLAY IRFU, CEA, Universit\'{e} Paris-Saclay, F-91191 Gif-sur-Yvette, France}
\affiliation{\MSU Skobeltsyn Nuclear Physics Institute and Physics Department at Lomonosov Moscow State University, 119899 Moscow, Russia}
\affiliation{\SCAROLINA University of South Carolina, Columbia, South Carolina 29208}
\affiliation{\TEMPLE Temple University, Philadelphia, Pennsylvania 19122}
\affiliation{\UTFSM Universidad T\'{e}cnica Federico Santa Mar\'{i}a, Casilla 110-V Valpara\'{i}so, Chile}
\affiliation{\JLAB Thomas Jefferson National Accelerator Facility, Newport News, Virginia 23606}
\affiliation{\ULS Universidad de La Serena, Avda. Juan Cisternas 1200, La Serena, Chile}
\affiliation{\YEREVAN Yerevan Physics Institute, 375036 Yerevan, Armenia}
\affiliation{\YORK University of York, York YO10 5DD, United Kingdom}

\date{\today}

\begin{abstract}
    Hyperon recoil polarization measurements for the exclusive electroproduction of $K^+\Lambda$ and $K^+\Sigma^0$ final states from an unpolarized 
    proton target have been carried out using the CLAS12 spectrometer at Jefferson Laboratory. The measurements at beam energies of 6.535~GeV and 
    7.546~GeV span the range of four-momentum transfer $Q^2$ from 0.3 to 4.5~GeV$^2$ and invariant mass $W$ from 1.6 to 2.4~GeV, while covering 
    the full center-of-mass angular range of the $K^+$. These new $\Lambda$ polarization observables extend the existing data in a similar kinematic 
    range but from a significantly larger dataset. However, they represent the first electroproduction measurements of this observable for the 
    $\Sigma^0$. These data will allow for better exploration of the reaction mechanism in strangeness production, for further understanding of the 
    spectrum and structure of excited nucleon states that couple to $KY$, and for improved insight into the strong interaction in the non-perturbative 
    domain.
\end{abstract}


\maketitle

\section{Introduction}
\label{intro}

In recent years the expanding database of observables from exclusive meson photo- and electroproduction experiments has allowed for significant 
advancement in mapping out the spectrum of the excited states of the nucleon ($N^*$s) and in understanding their internal structure. The spectrum 
and structure of these resonance states encode the dynamics of Quantum Chromodynamics (QCD) from the perturbative (pQCD) to the strongly coupled 
(strong QCD) regimes where confinement and dynamical hadron mass generation in connection with chiral symmetry breaking lead to the properties of 
matter. Studies of the strong interaction dynamics that govern the generation of hadron ground and excited states in the non-perturbative regime 
where the running coupling of QCD is large, {\it i.e.} $\alpha_s/\pi \approx 1$ (the strong QCD regime), represent a crucial challenge in modern 
hadron physics. See Refs.~\cite{Burkert:2019bhp,qcd2019,fbs-carman} for recent reviews of the field.

A comprehensive research program is underway in Hall~B at Jefferson Laboratory (JLab) with the objective to determine the nucleon resonance
electroexcitation amplitudes through different electroproduction reactions from an unpolarized proton target~\cite{fbs-carman,fbs-mokeev}. 
These $\gamma_v pN^*$ electrocouplings are unambiguously related to the $N \to N^*$ transition form factors. The program in Hall~B has
provided these amplitudes for most $N^*$ states in the mass range up to 1.8~GeV for photon virtualities $Q^2$ up to 5~GeV$^2$ based on data 
collected with the CLAS spectrometer~\cite{clas-nim} during the JLab 6-GeV era. These studies offer unique information on the strong QCD dynamics 
that govern the generation of these states with different quantum numbers and distinctively different structural properties. Studies of nucleon 
resonances created through multiple exclusive meson electroproduction channels are of particular importance for the extraction of the electrocouplings, 
as the ratio of resonant to non-resonant contributions varies significantly from channel-to-channel. However, the electrocouplings should be the same, 
since the resonance electroexcitation and hadronic decay amplitudes are independent. These studies of baryon structure are complementary to studies of 
meson structure, which together are of importance in order to understand the dynamics of the processes that generate the dominant portion of visible 
hadron mass in the Universe~\cite{roberts2020}.

The recent progress in understanding the spectrum and structure of $N^*$ states has mainly been provided by advanced analyses of the CLAS data 
for exclusive electroproduction of the $\pi N$ and $\pi^+ \pi^- p$ channels from a proton target. See Ref.~\cite{fbs-carman} for a recent review. 
However, the high-precision data from CLAS on exclusive photoproduction of $K^+Y$ ($Y=\Lambda,\Sigma^0$) 
\cite{mcnabb,bradford2006,bradford2007,mccracken2010,dey2010,paterson} have been crucial in this advancement, in particular, for the discovery of 
new baryon states previously known as ``missing" resonances. In the strangeness photoproduction channels, data are also available from MAMI
\cite{mami1}, SAPHIR~\cite{saphir1}, GRAAL~\cite{graal1,graal2}, LEPS~\cite{leps1,leps2}, and BGO-OD~\cite{bgood1}. The exclusive production of 
$KY$ is sensitive to coupling to higher-lying $N^*$ states for $W > 1.6$~GeV, which is precisely the mass range where the understanding of the $N^*$ 
spectrum is most limited. With these data, several $N^*$ states have recently been claimed within global multi-channel analyses of the exclusive 
photoproduction data with a decisive impact from the hyperon polarization observables~\cite{Bur17,Burkert:2020}. 

Electroproduction processes provide additional and complementary information to photoproduction as the photon virtuality $Q^2$ allows the spectrum
and structure of $N^*$ states to be probed as a function of the distance scale (or wavelength of the virtual photon probe). CLAS has provided most 
of the available world data results on cross sections~\cite{5st,carman13} and polarization observable~\cite{carman03,raue-car,sltp,carman09,ipol} 
for $K^+Y$ electroproduction in the nucleon resonance region. These measurements span $Q^2$ from 0.3 to 5.4~GeV$^2$, invariant mass $W$ from 1.6 to 
2.7~GeV, and cover the full center-of-mass (c.m.) angular range of the $K^+$. See Refs.~\cite{carman16,carman18} for reviews on the CLAS 
electroproduction datasets. In addition to the $KY$ electroproduction data from CLAS, measurements have also been provided by JLab Hall~A~\cite{Coman} 
and Hall~C~\cite{Mohring,Niculescu}, as well as by MAMI~\cite{Achenbach:2011rf}.

Gaining insight into QCD in the non-perturbative regime requires not only high-precision experimental data for both cross sections and polarization
observables but advanced reaction models that provide connections between the data and the $N/N^*$ structural parameters, such as the nucleon elastic 
form factors and $\gamma_vpN^*$ electrocouplings. These reaction models include single-channel models
\cite{fbs-mokeev,saclay-lyon1,saclay-lyon2,kaon-maid1,kaon-maid2,rpr,maxwell,az2013,skoupil18,skoupil24} and dynamical coupled-channel models
\cite{anl-osaka,jbw-model}. However, the constraints and insights that such phenomenological reaction models provide to understand the strong 
QCD dynamics are only as good as the quality of the experimental data. The less precise and complete the available data, the larger the uncertainties 
and ambiguities in the extracted multipoles and $\gamma_vpN^*$ electrocouplings that encode the nucleon resonance structure information.

Improving the statistical and systematic precision, as well as extending the kinematic range of the electroproduction data on the $KY$ differential 
cross sections and polarization observables, will be critical to foster these efforts. One of the goals of measuring $KY$ electroproduction with the 
new CLAS12 spectrometer in Hall~B at JLab~\cite{clas12-nim} is to provide data in the $Q^2$ range up to $\approx$5~GeV$^2$ at the same level of accuracy
as the available photoproduction data, while ultimately extending the extracted observables up to $Q^2$ of 10-12~GeV$^2$~\cite{e1216010a,e1206108a}. 

As a first step in this direction from the 12-GeV era at JLab, the beam-recoil transferred polarization for the $K^+\Lambda$ and $K^+\Sigma^0$ final 
states from the CLAS12 RG-K dataset was recently published~\cite{carman-tpol}, expanding and extending the results from the CLAS program
\cite{carman03,carman09}. This current work on the $\Lambda$ and $\Sigma^0$ recoil polarization is a continuation of these efforts to expand the 
available data from CLAS in order to provide more complete data as input to available reaction models to improve our understanding of strong QCD in 
the non-perturbative domain.

The $\Lambda$ recoil polarization in the exclusive $ep$ reaction was first measured from the CLAS e1f experiment using a 5.479~GeV electron beam
\cite{ipol}. The e1f dataset collected 5G triggers. The data spanned $Q^2$ from 0.8--3.5~GeV$^2$ and $W$ from 1.6--2.7~GeV. Due to limited statistics 
the polarization was provided in coarse bins in $W$ and $\cos \theta_K^{c.m.}$ integrated over $Q^2$. The recoil polarization of the ground state hyperons 
was also measured in the $\gamma p$ photoproduction reaction from the CLAS g1c and g11a datasets. The $\Lambda$ recoil polarization
\cite{mcnabb,mccracken2010} and $\Sigma^0$ recoil polarization~\cite{dey2010} were provided in fine binning for both $W$ and $\cos \theta_K^{c.m.}$ from 
the reaction threshold up to $W$=2.835~GeV. The number of triggers collected in these photoproduction datasets was roughly 5 times more than the CLAS e1f
electroproduction dataset.

This analysis of the recoil hyperon polarization for the $K^+\Lambda$ and $K^+\Sigma^0$ final states from the CLAS12 RG-K dataset represents a companion 
analysis to the beam-recoil hyperon polarizations measured from this same dataset. Essentially the same analysis code and event selection cuts were 
employed, the same binning was chosen, and the yield fitting algorithms were the same. The hyperon polarization transfer data from Ref.~\cite{carman-tpol} 
was based on the initial data processing that has since been enhanced with improved forward tracking efficiency due to the use of artificial intelligence 
algorithms for cluster finding and denoising that resulted in efficiency improvements of $\sim$10\% per charged particle track in the CLAS12 Forward 
Detector~\cite{gagik2024,denoise}. Additionally, an improved central tracker internal alignment procedure was instituted that resulted in significantly 
improved momentum resolution ($\Delta p/p \sim 5\%$ vs. 15\%) in the CLAS12 Central Detector~\cite{sebouh}.

The data from RG-K have roughly 10 times the statistics for the $K^+\Lambda$ analysis compared to the available CLAS e1f electroproduction dataset and 
span a similar kinematic regime. The data included here for the recoil polarization of the $\Sigma^0$ in the exclusive $e'K^+\Sigma^0$ final state 
represent the first time this observable has become available in electroproduction. Note that although the two ground-state hyperons have the same $uds$ 
valence quark content, their isospins differ ($I$=0 for $\Lambda$ and $I$=1 for $\Sigma^0$), so that $N^*$ states of $I=1/2$ can decay to $K^+\Lambda$, 
but $\Delta^*$ states cannot. Since both $N^*$ and $\Delta^*$ resonances can couple to the $K^+\Sigma^0$ final state, the hyperon final state selection 
amounts to an effective isospin filter.

The organization for the remainder of this paper is as follows. In Section~\ref{formalism} the formalism of the recoil polarization and the polarization 
extraction approach are presented. Section~\ref{expt} provides details on the RG-K experiment and data analysis procedures. Section~\ref{monte} provides 
an overview of the Monte Carlo (MC) used for the yield extraction and the acceptance function determination. Section~\ref{yields} details the yield 
extraction procedure. A discussion of the sources of systematic uncertainty is provided in Section~\ref{systematics}. Section~\ref{results} presents the 
measured $\Lambda$ and $\Sigma^0$ recoil polarizations from the CLAS12 data compared with several model predictions. Finally, a summary of this work and 
our conclusions are given in Section~\ref{conclusions}.

\section{Formalism and Polarization Extraction}
\label{formalism}

\subsection{Formalism}

\begin{figure}[htbp]
\centering
\includegraphics[width=0.9\columnwidth]{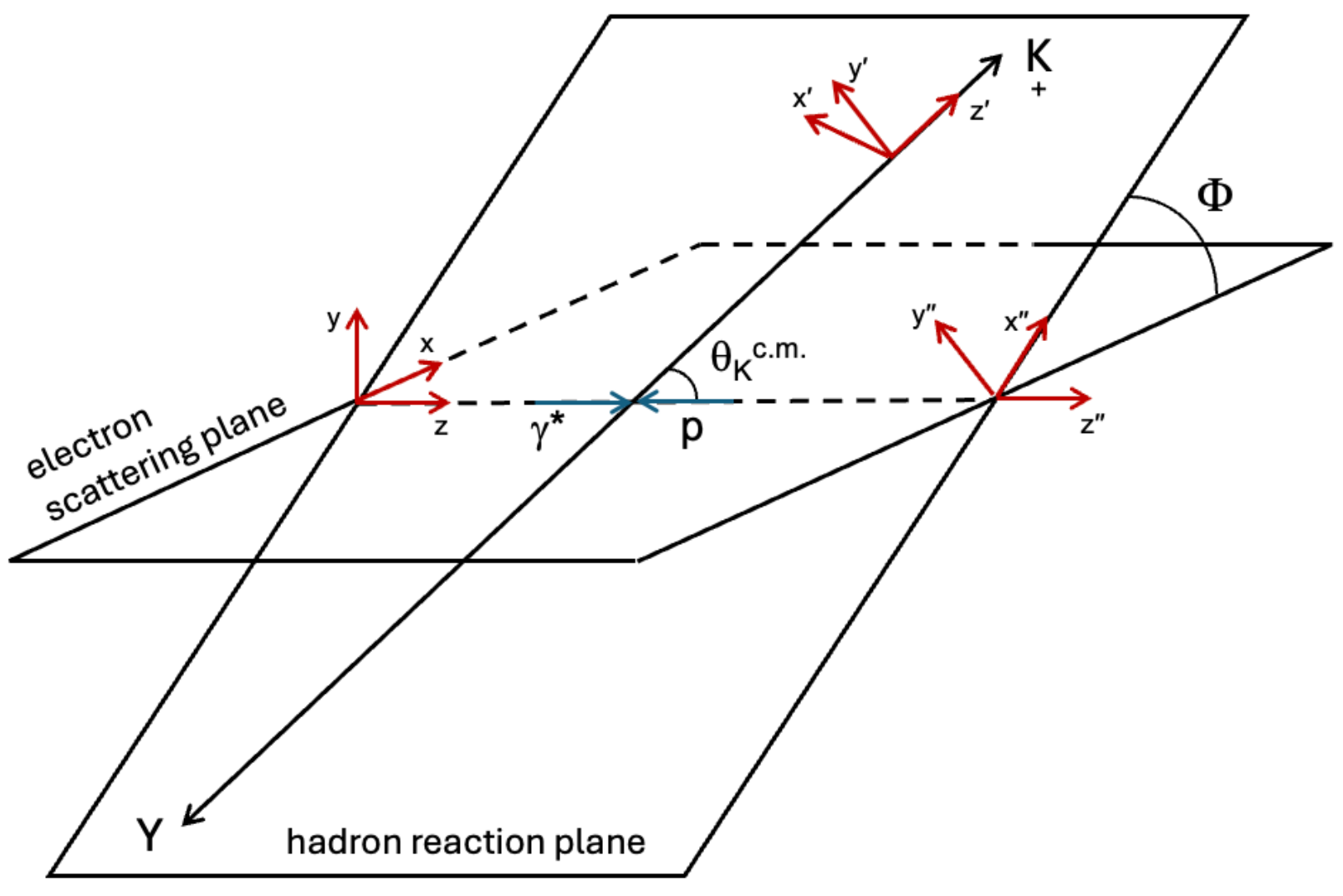}
\caption{Kinematics for $K^+Y$ electroproduction defining the c.m. angles and coordinate systems used to express the formalism and 
to present the polarization components extracted in the analysis.}
\label{coor4}
\end{figure}

For $KY$ electroproduction, the reaction kinematics are uniquely defined by the set of four variables ($Q^2$, $W$, $\cos \theta_K^{c.m.}$, $\Phi$), where 
$\theta_K^{c.m.}$ is the kaon production angle in the virtual photon-proton center-of-mass (c.m.) frame defined in Fig.~\ref{coor4} and $\Phi$ is the 
angle between the electron-scattering and the hadron-production planes. The squared four-momentum transfer of the virtual photon to the unpolarized target 
proton is $Q^2=-q^2$ and $W=\sqrt{M_p^2+2M_p\nu-Q^2}$ is the invariant mass of the final hadron state or the initial virtual photon-proton state, 
where $M_p$ is the proton mass and $\nu=E_e-E_{e'}$ is the virtual photon energy or the difference between the incident and scattered electron energies, 
respectively, in the laboratory frame.

The differential cross section for kaon electroproduction can be written as the product of a virtual photon flux factor $\Gamma_v$ and the kaon virtual 
photo-absorption cross section expressed in the kaon c.m. frame in the single photon exchange approximation as

\begin{equation}
\frac{d^5\sigma}{d\Omega_{e'} d\Omega_K^{c.m.} dE_{e'}} = \Gamma_v \frac{d\sigma_v}{d\Omega_K^{c.m.}}.
\end{equation}

\noindent
Following the notation of Ref.~\cite{knochlein}, the most general form for the photo-absorption cross section of a kaon from a proton target, allowing 
for a polarized electron beam, target proton, and recoil hyperon, is given by

\begin{widetext}
\begin{eqnarray}
\label{csec1}
\frac{d\sigma_v}{d\Omega_K^{c.m.}} &=& K_f \sum_{\alpha,\beta} S_\alpha S_\beta \Bigl[ R_T^{\beta\alpha} + \epsilon R_L^{\beta\alpha} 
+ \sqrt{\epsilon (1+\epsilon)} ( ^c\!R_{LT}^{\beta\alpha} \cos{\Phi} + \!^s\!R_{LT}^{\beta\alpha} \sin{\Phi}) \nonumber \\
&+& \epsilon(^c\!R_{TT}^{\beta\alpha} \cos{2\Phi} + \!^s\!R_{TT}^{\beta\alpha} \sin{2\Phi})
+ h\sqrt{\epsilon (1-\epsilon)} (^c\!R_{LT'}^{\beta\alpha} \cos{\Phi} + \!^s\!R_{LT'}^{\beta\alpha}
  \sin{\Phi}) + h \sqrt{1-\epsilon^2} R_{TT'}^{\beta\alpha} \Bigr].
\end{eqnarray}
\end{widetext}

\noindent
In this expression, the $R^{\beta \alpha}$ terms represent the response functions that account for the structure of the hadronic system. These response 
functions are themselves functions of $Q^2$, $W$, and $\cos \theta_K^{c.m.}$ (which is why these are the relevant kinematic variables for this analysis). 
Table~I of Ref.~\cite{ipol} lists the non-vanishing response functions for pseudoscalar meson electroproduction. Here $\epsilon$ is the virtual photon 
polarization parameter, $h$ is the electron-beam helicity, $K_f$ is a kinematic factor given by the ratio of the c.m. momenta of the outgoing kaon and 
the virtual photon, and the superscripts $\alpha$ and $\beta$ refer to the target and hyperon polarization coordinate systems, respectively. The $c$ or 
$s$ superscripts on the response function terms indicate a cosine or sine dependence of the respective contribution on the angle $\Phi$. Finally, 
$S_\alpha$ and $S_\beta$ are polarization-projection operators that serve to project the target and hyperon polarization vectors onto the different spin
quantization axes in the $(x,y,z)$ system for the target and the $(x',y',z')$ system for the hyperon (see the coordinate system definitions in 
Fig.~\ref{coor4}).

In the case where there is no beam, target, or recoil polarization, Eq.(\ref{csec1}) reduces to 

\begin{widetext}
\begin{equation}
\label{csec2}
\sigma_0 \equiv \left( \frac{d\sigma_v}{d\Omega_K^{c.m.}} \right)^{\!\!00} = K_f \left[ R_T^{00} +
  \epsilon R_L^{00} +\sqrt{\epsilon (1+\epsilon)} {^c\!R_{LT}^{00}} \cos{\Phi} + \epsilon {^c\!R_{TT}^{00}}
  \cos{2\Phi} \right],
\end{equation}
\end{widetext}

\noindent
so that $K_f R_i^{00} = \sigma_i$ of the usual unpolarized cross section notation with $i = L, T, LT, TT$. If we consider the general case of a polarized 
electron beam incident on an unpolarized target producing a polarized recoil hyperon, Eq.(\ref{csec1}) reduces to

\begin{equation}
\label{csec3}
\frac{d\sigma_v}{d\Omega_K^{c.m.}} = \sigma_0 (1 + h A_{LT'} +  P_{x'} \hat{x}' \cdot \hat{S}' +
P_{y'} \hat{y}' \cdot \hat{S}' + P_{z'} \hat{z}' \cdot \hat{S}'),
\end{equation}

\noindent
where $A_{LT'}$ is the polarized beam asymmetry defined in terms of the response function $R_{LT'}^{00}$.

Each of the hyperon polarization terms in Eq.(\ref{csec3}), $P_{x'}$, $P_{y'}$, $P_{z'}$, can be split into a beam helicity independent part, called the 
{\em recoil} or {\em induced} polarization, and a beam helicity dependent part, called the {\em transferred} polarization. The polarization components can 
be written as

\begin{equation}
\label{pol_def}
P_{i'} = P_{i'}^0 + h P_{i'}'.
\end{equation}

The three recoil polarization components are written in terms of the response functions in the $(x',y',z')$ system as~\cite{knochlein}

\begin{widetext}
\begin{eqnarray}
\label{pind1}
P_{x'}^0 &=& \frac{K_f}{\sigma_0} \left( \sqrt{\epsilon (1+\epsilon)} 
{^s\!R_{LT}^{x'0}}\sin{\Phi} + \epsilon {^s\!R_{TT}^{x'0}}\sin{2\Phi} \right) \nonumber\\
P_{y'}^0 &=& \frac{K_f}{\sigma_0} \left( R_T^{y'0} + \epsilon R_L^{y'0}
+ \sqrt{\epsilon (1+\epsilon)} {^c\!R_{LT}^{y'0}} \cos{\Phi}
+ \epsilon {^c\!R_{TT}^{y'0}} \cos{2\Phi} \right) \nonumber \\
P_{z'}^0 &=& \frac{K_f}{\sigma_0} \left( \sqrt{\epsilon (1+\epsilon)} 
{^s\!R_{LT}^{z'0}} \sin{\Phi} + \epsilon {^s\!R_{TT}^{z'0}} \sin{2\Phi} \right).
\end{eqnarray}
\end{widetext}

To accommodate finite bin sizes and to improve statistics, this analysis was performed by summing over all $\Phi$ angles. In this case, Eq.(\ref{pind1}) 
needs to be integrated over $\Phi$, and as a result, only the $y'$ component of the recoil polarization normal to the hadronic reaction plane is non-zero,

\begin{equation}
{\cal P}^0_{y'} = \frac{R_T^{y'0} + \epsilon R_L^{y'0}}{R_T^{00} + \epsilon R_L^{00}}.
\end{equation}

\noindent
Here the script symbol ${\cal P}^0$ denotes the $\Phi$-integrated polarization component.

%%%%%%%%%%%%%%%%%%%%%%%%%%%%%%%%%%%%%%%%%%%%%%%%%%%%%%%%%%%%%%%%%%%%%%%%%%%%%%%%%%%%%%%%%%%%%%%%%%%%%%%%%%%%%%%%%%%%%%%%%%%%%%%%%%
%%%%%%%%%%%%%%%%%%%%%%%%%%%%%%%%%%%%%%%%%%%%%%%%%%%%%%%%%%%%%%%%%%%%%%%%%%%%%%%%%%%%%%%%%%%%%%%%%%%%%%%%%%%%%%%%%%%%%%%%%%%%%%%%%%

To define the recoil polarization components in the $(x,y,z)$ system, the components defined for the $(x',y',z')$ system in Eq.(\ref{pind1}) must undergo
a simple transformation that performs a rotation of $\theta_K^{c.m.}$ about the $y'$ axis followed by a rotation of $\Phi$ about the $z'$ axis. With the 
integration over $\Phi$ for the $(x,y,z)$ system, again only the normal component of the recoil polarization is non-zero, this time normal to the electron 
scattering plane, and is given by

\begin{equation}
{\cal P}^0_y = \frac{1}{2} \frac{\sqrt{\epsilon (1+\epsilon)}}{R_T^{00} + \epsilon R_L^{00}} 
({^s\!R_{LT}^{x'0}} \cos{\theta_K^{c.m.}} + {^c\!R_{LT}^{y'0}} + {^s\!R_{LT}^{z'0}} \sin{\theta_K^{c.m.}}).
\end{equation}

\noindent
The extraction of the polarization components in the different coordinate systems can be seen to give sensitivity to different sets of response 
functions. 

\subsection{Hyperon Polarization Extraction}
\label{extraction}

The $\Lambda$ decays into a pion and a nucleon via the weak nuclear force. This decay has an asymmetric angular distribution with respect to the 
$\Lambda$ spin direction that results from an interference between parity non-conserving ($s$-wave) and parity-conserving ($p$-wave) amplitudes. The 
$\Lambda$ decays via both $p \pi^-$ and $n\pi^0$ with branching ratios of 63.9\% and 35.8\%, respectively. In this analysis, only the charged $\Lambda$ 
decay branch is considered. In the $\Lambda$ decay frame (or rest frame), the angular distribution of the $\Lambda$ decay proton can be written as
\cite{fasano}

\begin{equation}
  \label{ang-dist1}
  \frac{dN}{d \cos \theta_p^{\Lambda RF}} = N_0 \left( 1 + \alpha P_\Lambda \cos \theta_p^{\Lambda RF} \right),
\end{equation}

\noindent
where $N_0$ is the yield integral, $\alpha = 0.747 \pm 0.009$ is the $\Lambda$ weak decay asymmetry parameter~\cite{pdg}, $P_\Lambda$ is the 
$\Lambda$ polarization, and $\theta_p^{\Lambda RF}$ is the angle of the $\Lambda$ decay proton relative to a chosen spin quantization axis in the 
$\Lambda$ decay frame, {\it i.e.} its angle with respect to either the $y'$ or $y$ axis for the case of the recoil polarization measurements shown 
in this work.

The form of Eq.(\ref{ang-dist1}) connects the $\Lambda$ polarization to the angular distribution of its decay products and is the reason why the 
$\Lambda$ is said to be ``self-analyzing'' in its decay. This expression is also equally valid to measure the polarization of the $\Sigma^0$ via its 
electromagnetic decay into $\Lambda + \gamma$ (with branching ratio 100\%). To realize an expression appropriate for the $\Sigma^0$, two important 
polarization ``dilution'' effects must be taken into account. The polarization transfer from the $\Sigma^0$ to the $\Lambda$ is given by

\begin{equation}
  \label{spintrans}
  P_\Lambda = - P_\Sigma \cos \theta_\Lambda^{\Sigma RF},
\end{equation}

\noindent
where $\theta_\Lambda^{\Sigma RF}$ is the angle of the $\Lambda$ in the $\Sigma^0$ decay frame with respect to a given spin quantization axis. This 
relationship arises from evaluating the expectation value of the spin operator of the daughter $\Lambda$ in terms of the transition matrix for the 
electromagnetic decay of the parent $\Sigma^0$. This expression is derived in Ref.~\cite{bradford2007}. Note that in terms of spin structure, the 
$\Sigma^0 \to \Lambda \gamma$ reaction is a $\frac{1}{2} \to \frac{1}{2} \oplus 1$ decay, while Eq.(\ref{spintrans}) is obtained after averaging over 
the spin projections of the unobserved outgoing photon. Therefore, there is a dilution of the accessible $\Sigma^0$ spin information in this decay 
given by

\begin{equation}
  P_\Lambda = -\frac{1}{3} P_\Sigma.
\end{equation}

\noindent
Thus the magnitude of the average daughter $\Lambda$ polarization is 3 times smaller than that of the parent 
$\Sigma^0$ when averaging over all $\Lambda$ emission angles.

When Eq.(\ref{ang-dist1}) is combined with Eq.(\ref{spintrans}), the angular distribution of the decay proton relevant to access the $\Sigma^0$ 
polarization is given by

\begin{equation}
  \label{ang-dist2}
  \frac{dN}{d \cos \theta_p^{\Lambda RF}} = N_0 \left( 1 - \alpha P_\Sigma \cos \theta_\Lambda^{\Sigma RF}
  \cos \theta_p^{\Lambda RF} \right).
\end{equation}

\noindent
Note that Eq.(\ref{ang-dist2}) is written in terms of the proton angular distribution in the $\Lambda$ decay frame. However, as the $\Lambda$ momentum 
for the $K^+\Sigma^0$ final state is not accessible measuring only the $e'K^+p$ final state ({\it i.e.} the photon from the $\Sigma^0$ decay is not 
detected), the decay proton is actually measured in the parent $\Sigma^0$ decay frame. Thus, in addition to the dilution for the $\Sigma^0$ in the 
$\Lambda \gamma$ decay, a further dilution in the $\Sigma^0$ polarization occurs when measuring the decay proton in the $\Sigma^0$ decay frame. If the 
$\Sigma^0-\Lambda$ spin-transfer information is averaged over, then Eq.(\ref{ang-dist2}) is replaced by

\begin{equation}
\frac{dN}{d \cos \theta_p^{\Sigma RF}} = N_0 \left( 1 - \nu_\Sigma \alpha P_\Sigma \cos \theta_p^{\Sigma RF} \right),
\end{equation}

\noindent
where in this expression the angular distribution of the decay proton from the daughter $\Lambda$ is measured in the $\Sigma^0$ decay frame. The 
initial dilution factor of $\nu_\Sigma=-1/3$ now becomes $\nu_\Sigma=-1/3.90$ as shown in Appendix~A of Ref.~\cite{bradford2007}.

In summary, the general expression connecting the angular distribution of the hyperon decay proton to the polarization of either the parent $\Lambda$ 
or parent $\Sigma^0$ in the hyperon decay frame can be written as

\begin{equation}
  \label{ang-dist3}
  \frac{dN}{d \cos \theta_p^{RF}} = N_0 \left( 1 + \nu_Y \alpha P_Y \cos \theta_p^{RF} \right).
\end{equation}

\noindent
For the $K^+\Lambda$ analysis $\nu_\Lambda=+1.0$ and for the $K^+\Sigma^0$ analysis $\nu_\Sigma=-0.256$. It is assumed in this notation that the 
decay proton angle $\theta_p^{RF}$ is measured in the decay frame of the parent hyperon, either the $\Lambda$ or $\Sigma^0$, depending on the final 
state of interest.

The hyperon polarization vector $P_Y$ has components that can be computed with respect to a given set of spin-quantization axes ({\it i.e.} within a 
given coordinate system). For the general case, the hyperon polarization can be written as $P_Y = P^0_Y + P_b P'_Y$, where $P^0_Y$ is the recoil 
polarization and $P'_Y$ is the beam-recoil transferred polarization. Here, $P_b$ represents the longitudinal polarization of the incident beam electron.
In this analysis of the recoil polarization the data events with beam helicity $h=+1$ and $h=-1$ were summed together.

Relative to any particular choice of spin-quantization axis, the yield $N_F$ in the forward angle range (defined as $\cos \theta_p^{RF} > 0$) is
given by

\begin{eqnarray}
  N_F &=& \int_{0}^{1} N_0 \left( 1 + \nu_Y \alpha P_Y \cos \theta_p^{RF} \right) d\cos \theta_p^{RF}\\
  &=& N_0 + N_0 \cdot \frac{\nu_Y \alpha P_Y}{2}.
\end{eqnarray}

\noindent
Similarly, the yield $N_B$ in the backward angle range (defined as $\cos \theta_p^{RF} < 0$) is given by

\begin{eqnarray}
  N_B &=& \int_{-1}^{0} N_0 \left( 1 + \nu_Y \alpha P_Y \cos \theta_p^{RF} \right) d\cos \theta_p^{RF}\\
  &=& N_0 - N_0 \cdot \frac{\nu_Y \alpha P_Y}{2}.
\end{eqnarray}

The recoil polarization can be extracted from the forward-backward yield asymmetry $A_{FB}$ using

\begin{equation}
\label{afb1}
  {\cal P}^0_Y = \frac{2}{\nu_Y \alpha} \cdot A_{FB} = \frac{2}{\nu_Y \alpha} \cdot \cfrac{\frac{N_F}{A_F} - \frac{N_B}{A_B}}{\frac{N_F}{A_F} 
  + \frac{N_B}{A_B}},
\end{equation}

\noindent
where the relevant yields are properly acceptance-corrected using the appropriate forward acceptance $A_F$ and backward acceptance $A_B$ relevant for 
the given spin quantization axis. Note that the analysis of the beam-recoil transferred polarization is based on forming a helicity asymmetry
\cite{carman-tpol} that is essentially insensitive to detector acceptance effects. This analysis of the recoil polarization, however, is extremely 
sensitive to the affects of acceptance on the computed $A_{FB}$ asymmetries. See Section~\ref{monte} for more details on the affect of the acceptance 
on $A_{FB}$ for the $y'$ and $y$ axes.

\section{Experiment Description and Data Analysis}
\label{expt}

\subsection{CLAS12 Spectrometer}

\begin{figure}[htbp]
\centering
\includegraphics[width=1.0\columnwidth]{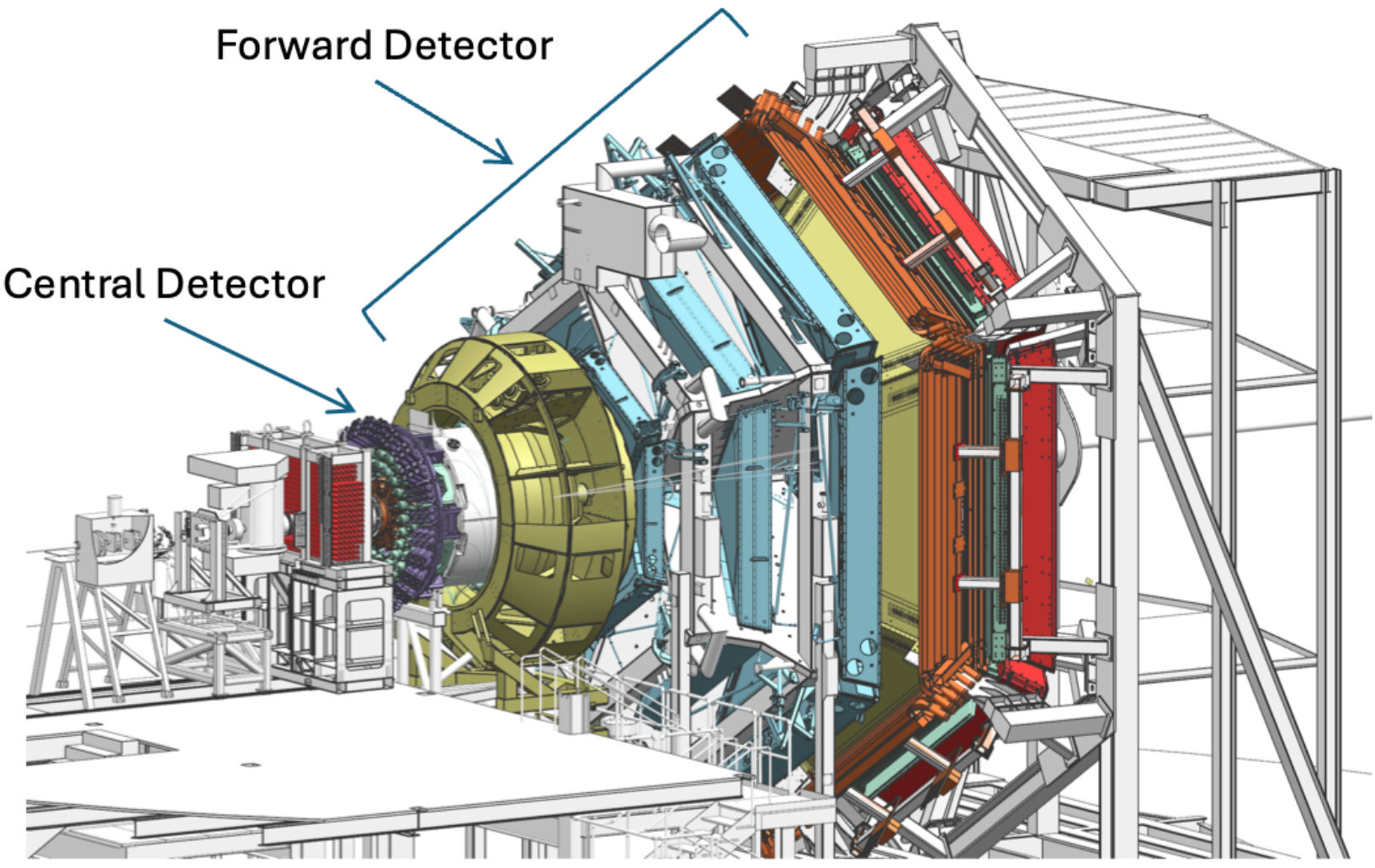}
\caption{Model of the CLAS12 spectrometer in Hall~B at Jefferson Laboratory. The electron beam is incident from the left side of this
  figure. The CLAS12 detector is roughly 20~m in scale along the beam axis. See Ref.~\cite{clas12-nim} for details.}
\label{clas12-model}
\end{figure}

The CLAS detector in Hall~B \cite{clas-nim} was replaced with the large acceptance CLAS12 spectrometer~\cite{clas12-nim} as part of the JLab 12~GeV 
upgrade project with the physics program beginning in 2018. CLAS12 consists of a Forward Detector system built around a superconducting torus magnet 
that divides the acceptance into six 60$^\circ$-wide azimuthal sectors and a Central Detector built around a superconducting solenoid magnet. 
Figure~\ref{clas12-model} shows a model representation of CLAS12. The Forward Detector covers polar angles $\theta$ from 5$^\circ$--35$^\circ$ and 
the Central Detector covers polar angles from 35$^\circ$--125$^\circ$. In the forward direction, CLAS12 consists of multiple sets of drift chambers
\cite{dc-nim} for charged particle tracking installed before, within, and after the torus field. Downstream of the drift chambers, the spectrometer 
includes a forward time-of-flight system for precise timing measurements for charged particles~\cite{ftof-nim} and a sampling electromagnetic 
calorimeter for identification of electrons, photons, and neutrons~\cite{ecal-nim}. The Forward Detector also includes a high threshold Cherenkov 
detector upstream of the drift chambers that is used as part of the electron trigger~\cite{htcc-nim}. The Central Detector consists of a multi-layer 
charged particle tracking system surrounding the target and beamline~\cite{svt-nim,mm-nim} and a central time-of-flight system for charged particle
identification~\cite{ctof-nim} via precise flight time measurements. Note that the brief description given here of CLAS12 only details the detector
subsystems used for this measurement of hyperon recoil polarization. CLAS12 includes other subsystems that are discussed in Ref.~\cite{clas12-nim}.

%%%%%%%%%%%%%%%%%%%%%%%%%%%%%%%%%%%%%%%%%%%%%%%%%%%%%%%%%%%%%%%%%%%%%%%%%%%%%%%%%%%%%%%%%%%%%%%%%%%%%%%%%%%%%%%%%%%%%%%%%%%%%%%%%%%%%%%%
%%%%%%%%%%%%%%%%%%%%%%%%%%%%%%%%%%%%%%%%%%%%%%%%%%%%%%%%%%%%%%%%%%%%%%%%%%%%%%%%%%%%%%%%%%%%%%%%%%%%%%%%%%%%%%%%%%%%%%%%%%%%%%%%%%%%%%%%

The data employed for this work were collected as part of the RG-K experiment that took place at the end of 2018. Data were taken at beam energies 
of 6.535~GeV and 7.546~GeV using a 5-cm-long liquid-hydrogen target. The 6.535~GeV (7.546~GeV) dataset was collected at an average beam-target 
luminosity of 1$\times$10$^{35}$~cm$^{-2}$s$^{-1}$ (5$\times$10$^{34}$~cm$^{-2}$s$^{-1}$) and amounted to 18.2~mC (10.7~mC) of accumulated electron 
charge. The electron beam was longitudinally polarized with an average beam polarization of 86\%. The torus magnet was set to its maximum field 
strength to optimize the reconstructed momentum resolution for charged particles and its polarity was set to bend negatively charged particles 
outward, away from the beamline. Event readout was triggered by a coincidence between a track candidate in the drift chamber, a signal in the high 
threshold Cherenkov detector, and a minimum-energy cluster in the forward calorimeter~\cite{trigger-nim}. The data acquisition system~\cite{daq-nim} 
recorded data at typical rates of 20~kHz and 500~MB/s with a live-time greater than 90\%.

\subsection{Analysis Cut Overview and Details}
\label{cuts}

\begin{figure}[htbp]
\centering
\includegraphics[width=1.0\columnwidth]{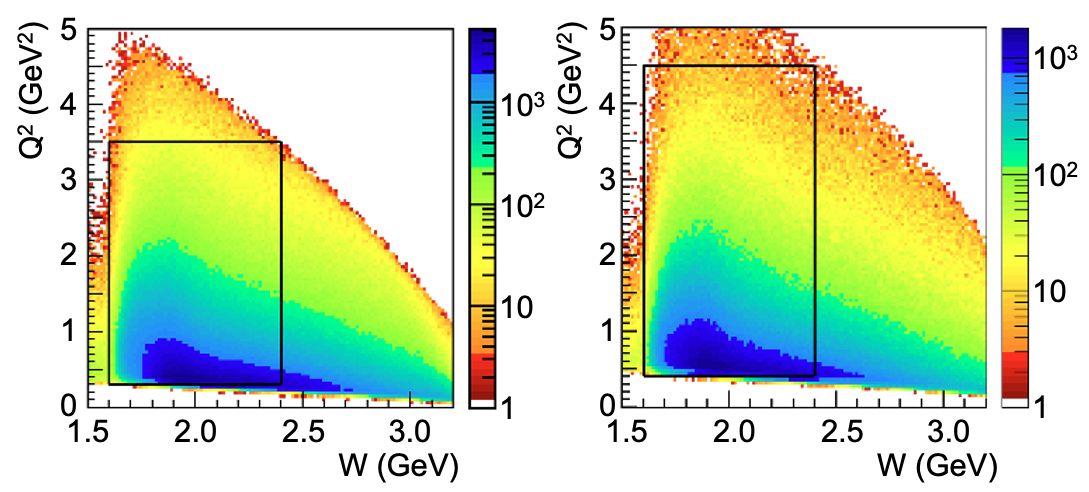}
\caption{Kinematics for $K^+Y$ electroproduction in terms of $Q^2$ (GeV$^2$) vs. $W$ (GeV) for the 6.535~GeV data (left) and the 
7.546~GeV data (right). This analysis focuses on the nucleon resonance region with $W$ from 1.6 -- 2.4~GeV and $Q^2$ from 0.3 --
3.5~GeV$^2$ for the 6.535~GeV dataset and $Q^2$ from 0.4 -- 4.5~GeV$^2$ for the 7.546~GeV dataset (represented by the rectangle 
drawn on each plot).}
\label{q2w}
\end{figure}

\begin{figure}[htbp]
\centering
\includegraphics[width=0.75\columnwidth]{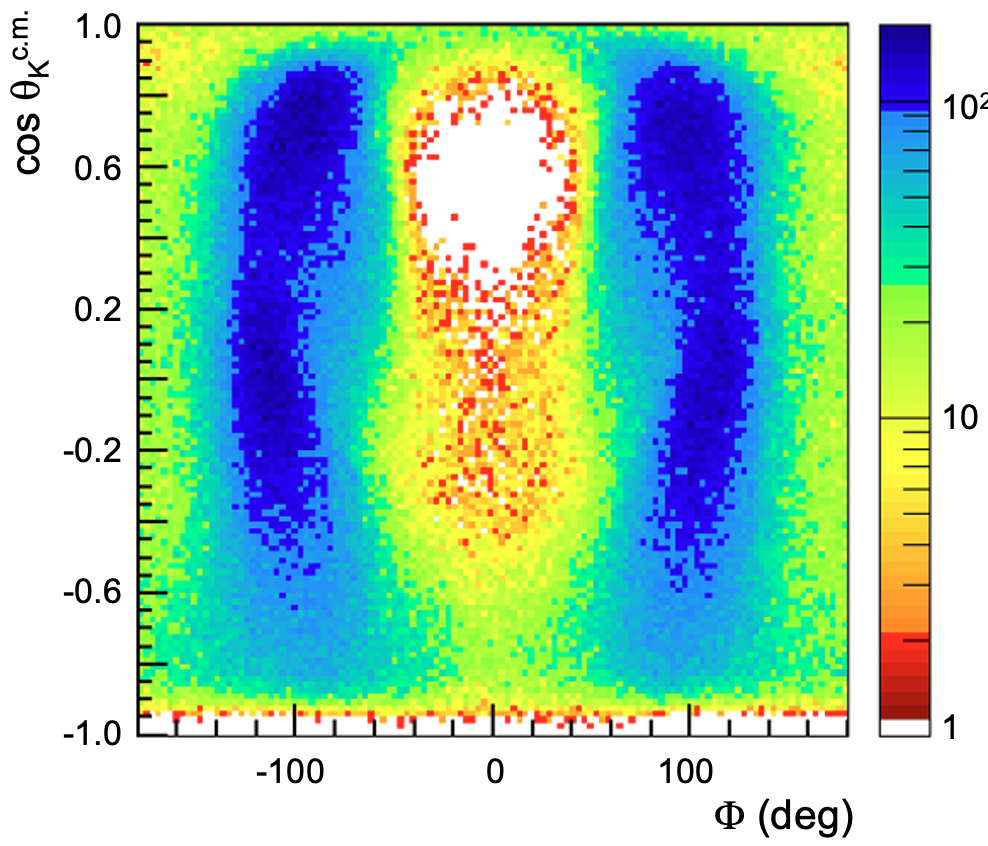}
\caption{Kinematic coverage of the electroproduced $K^+$ in terms of $\cos \theta_K^{c.m.}$ vs. $\Phi$ (deg), where $\Phi$ is the angle 
between the lepton scattering plane and the hadronic reaction plane for the 6.535~GeV dataset.}
\label{thphi}
\end{figure}

Hyperon identification relies on missing-mass reconstruction of the reaction $ep\!\to\!e'K^+Y$. In addition, for the polarization measurement, the 
reconstruction of the proton from the hyperon decay is required. The acceptance for this three-body $e'K^+p$ final state is on the order of 1\% to 6\% 
depending on $Q^2$, $W$, $\cos \theta_K^{c.m.}$, and $\cos \theta_p^{RF}$. Figure~\ref{q2w} shows the electron kinematic coverage of the datasets in 
terms of $Q^2$ vs.~$W$. Figure~\ref{thphi} shows the kinematic phase space for the electroproduced $K^+$ from the 6.535~GeV data in terms of 
$\cos \theta_K^{c.m.}$ vs. $\Phi$.

In the kinematic region of interest, the 6.535~GeV (7.546~GeV) dataset contains 980k (429k) $K^+\Lambda$ events and 445k (176k) $K^+\Sigma^0$ 
events in the $e'K^+p$ topology. This data sample is roughly 10 times larger than that for the polarization analyses of the available CLAS 
electroproduction datasets~\cite{carman03,carman09}. The data presented in this work represents only 10\% of the full approved data collection
totals for the RG-K experiment. In this section, details are provided on our procedures for particle identification, on the cuts used to isolate 
the $K^+\Lambda$ and $K^+\Sigma^0$ final states, and on other cuts and corrections that are part of the data analysis.

%%%%%%%%%%%%%%%%%%%%%%%%%%%%%%%%%%%%%%%%%%%%%%%%%%%%%%%%%%%%%%%%%%%%%%%%%%%%%%%%%%%%%%%%%%%%%%%%%%%%%%%%%%%%%%%%%%%%%%%%%%%%%%%%%%
%%%%%%%%%%%%%%%%%%%%%%%%%%%%%%%%%%%%%%%%%%%%%%%%%%%%%%%%%%%%%%%%%%%%%%%%%%%%%%%%%%%%%%%%%%%%%%%%%%%%%%%%%%%%%%%%%%%%%%%%%%%%%%%%%%

%%%%%%%%%%%%%%%%%%%%%%%%%%%%%%%%%%%%%%%%%%%%%%%%%%%%%%%%%%%%%%%%%%%%%%%%%%%%%%%%%%%%%%%%%%%%%%%%%%%%%%%%%%%%%%%%%%%%%%%%%%%%%%%%%%
%%%%%%%%%%%%%%%%%%%%%%%%%%%%%%%%%%%%%%%%%%%%%%%%%%%%%%%%%%%%%%%%%%%%%%%%%%%%%%%%%%%%%%%%%%%%%%%%%%%%%%%%%%%%%%%%%%%%%%%%%%%%%%%%%%

\subsubsection{Particle Identification}

Event reconstruction began by selecting events with an electron candidate in the CLAS12 Forward Detector. The initial identification of 
electrons was performed by the CLAS12 Event Builder~\cite{recon-nim}. This required a negatively charged particle - determined by the 
track curvature in the toroidal magnetic field - that was matched with hits in the high-threshold Cherenkov detector, forward time-of-flight 
system, and calorimeter. The deposited energy in the calorimeter was required to be consistent with the parameterized sampling fraction 
distribution vs.~deposited energy. This definition was already sufficient to remove the dominant pion contamination; however, the analysis 
applied additional cuts to further refine the electron sample. Cuts were placed on the electron momentum as reconstructed in the drift 
chamber system, the particle flight time through CLAS12, and the reconstructed event vertex coordinate to be sure the track originated 
from the target cell (the trace-back resolution at the target location is $\approx$5~mm). 

The hadron identification process searched within the selected event sample for events with one (and only one) reconstructed $K^+$ and $p$ 
candidate. The CLAS12 Event Builder algorithm for charged hadrons compared the measured flight time for each track from the event vertex to 
the time-of-flight systems, to the computed time for a given hadron species, starting from its measured momentum and the assumed mass. The 
hypothesis that minimized the time difference was assigned as the particle type ($\pi$, $K$, or $p$). Additional cuts were applied to improve 
the hadron identification purity on the minimum particle momentum (0.4~GeV in the Forward Detector and 0.2~GeV in the Central Detector) and 
the particle flight time in CLAS12.

\subsubsection{Additional Cuts and Corrections}
\label{add-cuts}

It is important to optimize the accuracy of the reconstructed momenta of the final state $e'$ and $K^+$ to maximize the hyperon 
signal-to-background ratio in the $MM(e'K^+)$ spectra and to enable the best possible separation of the $K^+\Lambda$ and $K^+\Sigma^0$ final 
states. It is also important to optimize the accuracy of momentum reconstruction of the final state particles ($e'$, $K^+$, and $p$) in order 
to minimize the systematic uncertainties of the measured proton angular distribution used to determine the hyperon polarization.

The measured charged particle momenta in CLAS12 have inaccuracies due to unaccounted for geometrical misalignments of the tracking detectors 
relative to the magnetic fields, systematic reconstruction and calibration biases, and inaccuracies in the magnetic field maps for the torus 
and solenoid used in the charged particle tracking. However, the systematics of the measured momenta from CLAS12 were minimized accounting
for the hadron energy loss through the passive detector materials using the CLAS12 MC (see Section~\ref{monte}) and correcting for momentum 
mismeasurements using kinematic constraints for the different final state particles based on exclusive event reconstruction of multiple final 
states ({\it e.g.} elastic $ep$, $\pi N$, and $K^+\Lambda$). In the Forward (Central) Detector of CLAS12 the momentum resolution 
$\Delta p/p \sim 0.5 - 1$\% (5\%) with average sector-dependent shifts $< 0.5$\% (0.5--1.0\%). After the energy loss and momentum corrections 
the residual distortions of the $MM(e'K^+)$ and $MM(e'K^+p)$ spectra were at a level below $\pm$5~MeV over the full kinematic phase space of 
the data. The remaining residual distortions of the reconstructed momenta were shown to have a minimal effect on the assigned systematic 
uncertainties of the extracted hyperon polarizations.

The reconstructed momentum of charged particles in the CLAS12 Forward Detector suffers from systematic inaccuracies at the boundaries of 
the azimuthal acceptance in each sector close to the torus coils. To remove these events, geometrical fiducial cuts were employed to
exclude tracks detected in these regions. For the electrons, a selection on the calorimeter fiducial volume was also applied to ensure
containment of the electromagnetic shower, such that the sampling fraction cuts allow for high purity of the electron candidate sample.

In the extraction of the hyperon polarization components no radiative corrections were applied to the data. The need for such corrections
was minimized by employing relatively strict hyperon selection cuts on the $MM(e'K^+)$ mass distributions to restrict the radiative tail 
events. With our relatively tight hyperon mass cuts, the maximum radiated photon energy is $< 35$~MeV, which has a negligible impact on 
our computed $\cos \theta_p^{RF}$ values with respect to each quantization axis.

\subsubsection{Final State Identification}
\label{hypid}

\begin{figure*}[htbp]
\centering
\includegraphics[width=0.90\textwidth]{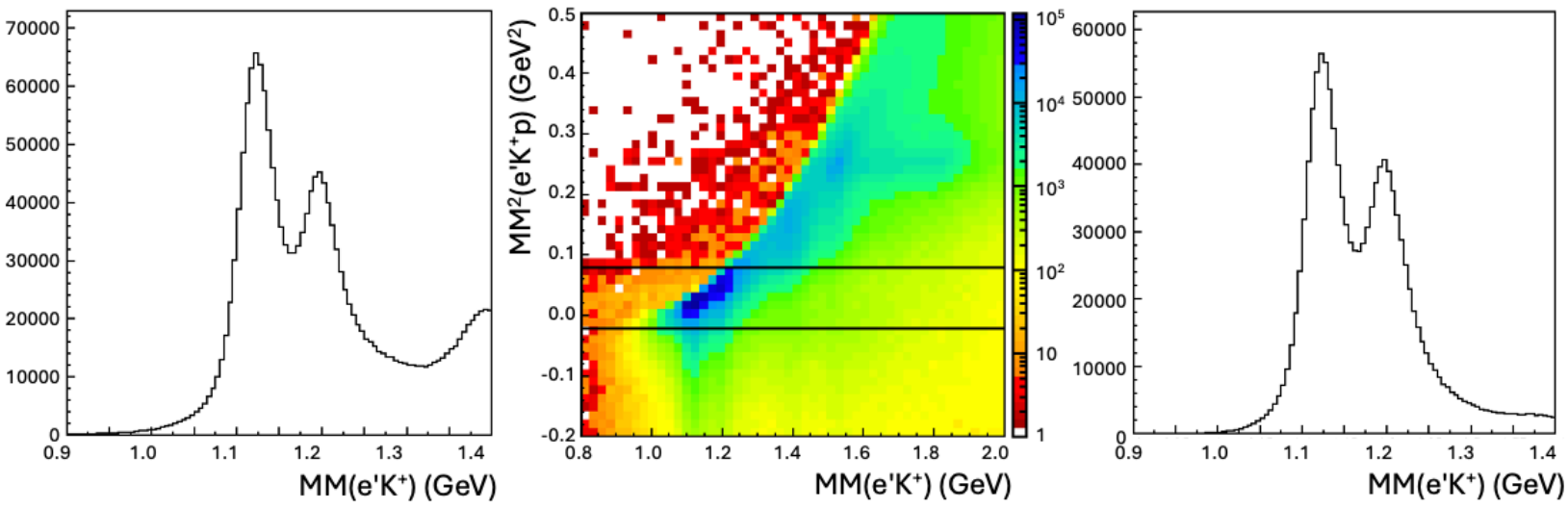}
\vspace{-3mm}
\caption{(Left) $MM(e'K^+)$ distribution requiring detection of a proton in the final state. (Middle) $MM^2(e'K^+p)$ vs.~$MM(e'K^+)$ phase 
space showing the cut employed on the $MM^2(e'K^+p)$ distribution to improve selection of the ground state hyperons. (Right) $MM(e'K^+)$ 
distribution shown in the left plot but with the cut on $MM^2(e'K^+p)$. Data are shown from the 6.535~GeV dataset.}
\label{hypcut}
\end{figure*}

The $K^+\Lambda$ and $K^+\Sigma^0$ final states were identified by selecting mass regions within the $MM(e'K^+)$ distribution. The backgrounds 
in these spectra were reduced using additional restrictions based on the reconstruction of the $e'K^+p$ final state. For $K^+\Lambda$ the 
missing particle is a $\pi^-$ and for $K^+\Sigma^0$ the missing particles are a $\pi^-$ and a low-momentum $\gamma$ from the $\Sigma^0$
electromagnetic decay. Cuts were applied on $MM^2(e'K^+p)$ from 0.00--0.08~GeV$^2$ to select the ground state hyperon region. 
Figure~\ref{hypcut} shows the $MM^2(e'K^+p)$ vs.~$MM(e'K^+)$ distribution phase space from the 6.535~GeV dataset, as well as the $MM(e'K^+)$ 
distribution before and after this additional cut. The $MM(e'K^+)$ spectrum before the cut shows an additional peak at about 1.4~GeV from the
contributions of the $\Sigma^0(1385)$ and $\Lambda(1405)$ hyperon excited states. The cut also serves to significantly reduce the background 
beneath the hyperon peaks that arises primary from the multi-pion channels with the $\pi^+$ misidentified as a $K^+$ due to the finite timing
resolutions of the CLAS12 time-of-flight systems. The hyperon $MM(e'K^+)$ resolution of CLAS12 for these RG-K datasets is 15-25~MeV depending 
on kinematics. It is relatively independent of $W$ and $\cos \theta_K^{c.m.}$ for the different $e'K^+$ topologies. However, it degrades slowly 
vs.~$Q^2$ from 15~MeV at $Q^2=0.3$~GeV$^2$ to 25~MeV at $Q^2=4.5$~GeV$^2$.

For this analysis, three different hadronic event topologies were combined together. The dominant topologies with roughly equal statistics
are $e'K^+_Fp_F$ and $e'K^+_Cp_F$, where the hadron subscripts $F$ and $C$ refer to whether the hadron was detected in the CLAS12 Forward
Detector or Central Detector, respectively. The $e'K^+_Fp_C$ topology contains only about 15\% of the event yield. The $e'K^+_Cp_C$ topology
is kinematically disfavored due to energy/momentum conservation with the electron detected in the forward direction and does not contribute
to the final event sample.

%%%%%%%%%%%%%%%%%%%%%%%%%%%%%%%%%%%%%%%%%%%%%%%%%%%%%%%%%%%%%%%%%%%%%%%%%%%%%%%%%%%%%%%%%%%%%%%%%%%%%%%%%%%%%%%%%%%%%%%%%%%%%%%%%%%%%%%%%%%
%%%%%%%%%%%%%%%%%%%%%%%%%%%%%%%%%%%%%%%%%%%%%%%%%%%%%%%%%%%%%%%%%%%%%%%%%%%%%%%%%%%%%%%%%%%%%%%%%%%%%%%%%%%%%%%%%%%%%%%%%%%%%%%%%%%%%%%%%%%

\subsection{Data Binning}
\label{binning}

The results shown in this work focus on the nucleon resonance region, spanning invariant mass $W$ from the $K^+Y$ threshold to 2.4~GeV. The 
recoil polarization components for the $K^+\Lambda$ and $K^+\Sigma^0$ final states are presented in a 1D binning scenario vs. $Q^2$, $W$, 
and $\cos \theta_K^{c.m.}$ integrated over the other two variables. The observables are also presented in a 3D binning scenario divided into 
2 bins in $Q^2$ and 4 bins of $\cos \theta_K^{c.m.}$. In this multi-dimensional binning the polarization observables are shown as a function 
of $W$. Table~\ref{1d-bin} presents the 1D binning choices and Table~\ref{3d-bin} presents the 3D binning choices for the 6.535~GeV and
7.546~GeV datasets. This binning choice matches that employed for the hyperon beam-recoil transferred polarization analysis from this same
dataset of Ref.~\cite{carman-tpol}.

%%%%%%%%%%%%%%%%%%%%%%%%%%%%%%%%%%%%%%%%%%%%%%%%%%%%%%%%%%%%%%%%%%%%%%%%%%%%%%%%%%%%%%%%%%%%%%%%%%%%%%%%%%%%%%%%%%%%%%%%%%%%%%%%%%
\begin{table}[htbp]
\centering
\begin{tabular}{|c|c|c|} \hline
Dependence       & Range                   & Bin Size \\ \hline
$Q^2$            & $Q^2_{min}$ - 1.5~GeV$^2$ & 0.1~GeV$^2$ \\
                 & 1.5 - 2.5~GeV$^2$       & 0.2~GeV$^2$ \\
                 & 2.5 - 3.1~GeV$^2$       & 0.3~GeV$^2$ \\
                 & 3.1 - 3.5~GeV$^2$       & 0.4~GeV$^2$ \\
                 & 3.5 - 4.5~GeV$^2$       & 1.0~GeV$^2$ \\ \hline
$W$              & $W_{min}$ - 2.4~GeV      & 25~MeV \\ \hline
$\cos \theta_K^{c.m.}$ & -1 $\to$ 1 & 0.08 \\ \hline
\end{tabular}
\caption{Binning for the 1D polarization analysis vs. $Q^2$, $W$, and $\cos \theta_K^{c.m.}$. The kinematic range spans $Q^2$ from 
0.3--3.5~GeV$^2$ for the 6.535~GeV dataset and from 0.4--4.5~GeV$^2$ for the 7.546~GeV dataset. For both datasets $W$ spans from
1.625~GeV (1.725~GeV) to 2.4~GeV for the $K^+\Lambda$ ($K^+\Sigma^0$) final state.}
\label{1d-bin}
\end{table}
%%%%%%%%%%%%%%%%%%%%%%%%%%%%%%%%%%%%%%%%%%%%%%%%%%%%%%%%%%%%%%%%%%%%%%%%%%%%%%%%%%%%%%%%%%%%%%%%%%%%%%%%%%%%%%%%%%%%%%%%%%%%%%%%%%

%%%%%%%%%%%%%%%%%%%%%%%%%%%%%%%%%%%%%%%%%%%%%%%%%%%%%%%%%%%%%%%%%%%%%%%%%%%%%%%%%%%%%%%%%%%%%%%%%%%%%%%%%%%%%%%%%%%%%%%%%%%%%%%%%%
\begin{table}[htbp]
\centering
\begin{tabular}{|c|c|} \hline
  Variable  & Bin Choices \\ \hline
  $Q^2$  & 0.3 -- 0.9~GeV$^2$, 0.9 -- 3.5~GeV$^2$ (6.535~GeV) \\
        &  0.4 -- 1.0~GeV$^2$, 1.0 -- 4.5~GeV$^2$ (7.546~GeV)\\ \hline
  $W$    & $W_{min}$ - 2.4~GeV in 80~MeV bins \\ \hline
  $\cos \theta_K^{c.m.}$ & -1 $\to$ 1 in 0.5 bins \\ \hline
\end{tabular}
\caption{Binning for the 3D polarization analysis in $Q^2$, $W$, and $\cos \theta_K^{c.m.}$ for the 6.535~GeV and 7.546~GeV datasets, where 
$W_{min}$=1.625~GeV (1.725~GeV) for the $K^+\Lambda$ ($K^+\Sigma^0$) final state.}
\label{3d-bin}
\end{table}
%%%%%%%%%%%%%%%%%%%%%%%%%%%%%%%%%%%%%%%%%%%%%%%%%%%%%%%%%%%%%%%%%%%%%%%%%%%%%%%%%%%%%%%%%%%%%%%%%%%%%%%%%%%%%%%%%%%%%%%%%%%%%%%%%%

In this analysis the bin sizes were uniform in $W$ and $\cos \theta_K^{c.m.}$. However, the $Q^2$ bin sizes increase with increasing $Q^2$ to 
compensate for the rapid fall-off of the cross sections. The results for all polarization components are reported at the geometric center of 
the kinematic bins.

%%%%%%%%%%%%%%%%%%%%%%%%%%%%%%%%%%%%%%%%%%%%%%%%%%%%%%%%%%%%%%%%%%%%%%%%%%%%%%%%%%%%%%%%%%%%%%%%%%%%%%%%%%%%%%%%%%%%%%%%%%%%%%%%%%

\section{Monte Carlo}
\label{monte}

The MC employed for this analysis was based on the CLAS12 GEANT4 simulation software called GEMC~\cite{sim-nim}. The {\tt genKYandOnePion} 
event generator (EG) was employed for the generation of the $K^+\Lambda$ and $K^+\Sigma^0$ events. This generator, described in detail in
Ref.~\cite{genkyandonepion}, is based on fits of the existing CLAS $K^+Y$ electroproduction data with physically motivated extrapolations into 
the full CLAS12 kinematic phase space for beam energies up to 11~GeV. 

The generator allows for setting the length and placement of the liquid-hydrogen target cell along the electron beamline to match the
experimental setup in RG-K. Note that the EG generates the hyperon decay uniform in the $\Lambda$ and $\Sigma^0$ rest frames, so does 
not include a mechanism to allow non-zero hyperon polarization.

The MC for this analysis was employed for two primary purposes. The first was to compute the CLAS12 acceptance function to account for the geometrical 
acceptance of the detector and the particle reconstruction efficiency with a model that closely matches the CLAS12 design. This is important given the 
sensitivity of the extracted polarization components to the corrected yields of Eq.(\ref{afb1}). The second purpose for the MC was to use the generated 
lineshapes for the $K^+\Lambda$ and $K^+\Sigma^0$ simulations as templates for the yield extraction fits in bins of $Q^2$, $W$, $\cos \theta_K^{c.m.}$, 
and $\cos \theta_p^{RF}$ (detailed in Section~\ref{yields}). Again close matching of the simulation to the data is essential to minimize systematic 
uncertainties in the extraction of the corrected yields.

In this regard several key modifications were made to the EG and its output truth information toward this aim. These changes that are detailed in the 
following subsections include

\begin{itemize}
\item Proper modeling of $\Lambda$ decay length relative to the primary $ep$ vertex,
\item Full accounting for radiative effects,
\item Accounting for non-zero hyperon polarization in the EG model to match the data.
\end{itemize}

\subsection{Event Generator Updates}

\subsubsection{Hyperon Decay Length}
\label{eg-mod1}

The original EG generated the $\Lambda \to p \pi^-$ decay at the primary $ep$ vertex or alternatively passed the hyperon parent four-momenta to GEMC 
to properly handle the hyperon decay. The former case systematically distorts the acceptance determination and the latter case does not allow for 
the EG truth information for the daughters to be carried with the event. The EG was updated to generate the $\Lambda$ decay length with 
$c \tau_\Lambda$=7.89~cm~\cite{pdg}. This modification to the EG was important to minimize systematic effects on the acceptance-corrected $N_F$ and 
$N_B$ yields with respect to the different coordinate axes.

\subsubsection{Internal Radiative Effects}
\label{eg-mod2}

The second modification to the MC was to more properly include the full effects of electron radiation that generate a tail to higher $MM(e'K^+)$. The 
{\tt genKYandOnePion} EG does not include a model for radiative effects. The GEANT4 simulation only accounts for external radiative effects for the 
scattered electron as it passes through the detector volume from the $ep$ interaction point through CLAS12. Thus, the important modifications to the 
hyperon lineshapes from internal radiative effects are not properly accounted for using the EG and GEANT4. Underestimating radiative contributions 
gives rise to systematic uncertainties in the spectrum fits, especially considering the strength of the radiative tail of the $\Lambda$ events 
beneath the $\Sigma^0$ events in the $MM(e'K^+)$ spectrum. To better account for the full set of radiative effects on the hyperon lineshape, 
additional radiation contributions were included in an effective manner following the approach in Ref.~\cite{moTsaiRC} (Appendix A). There it is 
shown that the probability of soft photon emission for internal radiation follows a $1/E_\gamma$ probability ({\it i.e.} a bremsstrahlung distribution) 
due to the contributing flux density that weights the real photon matrix element. Using this as motivation, an additional radiation contribution was 
included to reduce the scattered electron energy $E_{e'}$ with a $1/E_\gamma$ probability distribution event-by-event. A single normalization scale factor 
in the weighting was fixed to give the best overall $\chi^2$ for the $MM(e'K^+)$ spectrum fits. Studies of the internal radiation modeling are included 
in the systematic uncertainty studies of Section~\ref{systematics}.

\subsubsection{Polarization Modeling in Simulation Sample}
\label{eg-mod3}

The initial ${\cal P}^0$ extraction for the $\Lambda$ and $\Sigma^0$ was based on the nominal EG with the flat proton distributions in the hyperon 
decay frame that resulted in ${\cal P}^0_{MC}=0$. The final modification to the generated event sample was to re-weight the generated events from 
the initial EG to model the recoil polarization seen in the data vs. $Q^2$, $W$, and $\cos \theta_K^{c.m.}$. The final results are based on 
acceptance corrections using the realistic EG model. The final systematic uncertainty assignment is based on variations of the realistic EG physics 
model as discussed in Section~\ref{acc_func}.

\subsection{Acceptance Tables}

Two different approaches to determine the CLAS12 acceptance tables were developed for this analysis. Here they are referred to as the 2D and the 
4D acceptance models. Their generation and application are quite different. As such, the comparison of the hyperon polarization results extracted 
from both approaches provides for stringent tests of these corrections.

In the 2D approach, the $F$ and $B$ acceptance tables relevant for each spin-quantization axis were computed from the ratio of the reconstructed 
to generated events, $N_{rec}/N_{gen}$, in each kinematic bin. These 2D tables are binned exactly as for the 1D data sorts detailed in 
Section~\ref{binning}

\begin{itemize}
\item $Q^2$ sort: 20 $Q^2$ bins, 2 $\cos \theta_p^{RF}$ bins,
\item $W$ sort: 31 $W$ bins, 2 $\cos \theta_p^{RF}$ bins,
\item $\cos \theta_K^{c.m.}$ sort: 25 $\cos \theta_K^{c.m.}$ bins, 2 $\cos \theta_p^{RF}$ bins.
\end{itemize}

\begin{figure}[htbp]
\centering
\includegraphics[width=0.95\columnwidth]{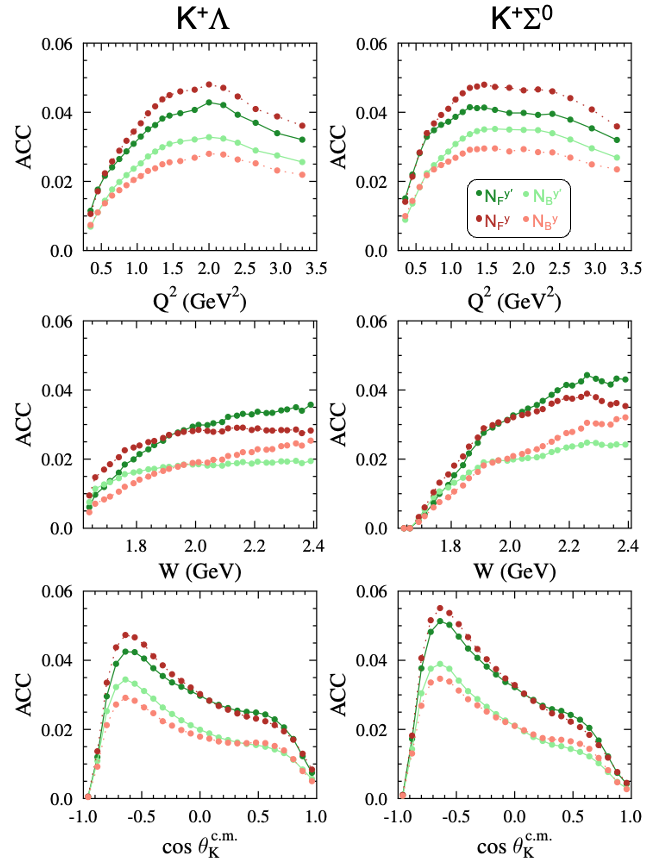}
\vspace{-4mm}
\caption{CLAS12 acceptance vs. $Q^2$ (top), $W$ (middle), and $\cos \theta_K^{c.m.}$ (bottom) summed over the other two variables for the 
$K^+\Lambda$ (left) and $K^+\Sigma^0$ (right) final states for $N_F$ and $N_B$ for the $y'$ and $y$ coordinates.}
\label{hyp-acc}
\end{figure}

Figure~\ref{hyp-acc} shows the derived acceptances for the $\Lambda$ and $\Sigma^0$ event samples, respectively, in the $e'K^+p$ final state for 
the 1D data sorts vs. $Q^2$, $W$, and $\cos \theta_K^{c.m.}$ from the 6.535~GeV MC. The acceptances are in the range from 1\% to 6\% for both 
hyperons.

%%%%%%%%%%%%%%%%%%%%%%%%%%%%%%%%%%%%%%%%%%%%%%%%%%%%%%%%%%%%%%%%%%%%%%%%%%%%%%%%%%%%%%%%%%%%%%%%%%%%%%%%%%%%%%%%%%%%%%%%%%%%%%%%%%
%%%%%%%%%%%%%%%%%%%%%%%%%%%%%%%%%%%%%%%%%%%%%%%%%%%%%%%%%%%%%%%%%%%%%%%%%%%%%%%%%%%%%%%%%%%%%%%%%%%%%%%%%%%%%%%%%%%%%%%%%%%%%%%%%%

The acceptance corrections in the 2D approach are applied to the binned yields $N_F$ and $N_B$ from fitting the binned $MM(e'K^+)$ 
distributions without acceptance corrections. Thus for each bin

\begin{equation}
A_{FB} = \frac{ \cfrac{\sum_i N_F}{ACC_{2D}} - \cfrac{\sum_i N_B}{ACC_{2D}} }
              { \cfrac{\sum_i N_F}{ACC_{2D}} + \cfrac{\sum_i N_B}{ACC_{2D}} },
\end{equation}

\noindent
where the sum runs over all $i$ events in the kinematic bin.

In the 4D approach the $F$ and $B$ acceptance tables were again computed as $N_{rec}/N_{gen}$ but with binning in $Q^2$, $W$, and 
$\cos \theta_K^{c.m.}$ as

\begin{itemize}
\item $Q^2$ sort: 20 $Q^2$ bins, 8 $W$ bins, 8 $\cos \theta_K^{c.m.}$ bins, 2 $\cos \theta_p^{RF}$ bins,
\item $W$ sort: 31 $W$ bins, 8 $Q^2$ bins, 8 $\cos \theta_K^{c.m.}$ bins, 2 $\cos \theta_p^{RF}$ bins,
\item $\cos \theta_K^{c.m.}$ sort: 25 $\cos \theta_K^{c.m.}$ bins, 8 $Q^2$ bins, 8 $W$ bins, 2 $\cos \theta_p^{RF}$ bins.
\end{itemize}
\noindent
The 4D acceptance tables were applied event-by-event during the data sorting, with the $\Lambda$ analysis run separately from the $\Sigma^0$
analysis. Thus

\begin{equation}
A_{FB} = \frac{\sum\limits_i \left( \cfrac{N_F}{ACC_{4D}} \right)_i - \sum\limits_i \left( \cfrac{N_B}{ACC_{4D}} \right)_i}
              {\sum\limits_i \left( \cfrac{N_F}{ACC_{4D}} \right)_i + \sum\limits_i \left( \cfrac{N_B}{ACC_{4D}} \right)_i}. 
\end{equation}

The systematics of the two approaches are distinctly different. The 2D approach, given binning only in one of the kinematic variables $Q^2$, $W$, 
or $\cos \theta_K^{c.m.}$, is more sensitive to bin migration effects arising from differences in the EG model and the data. The 4D approach uses 
the binned acceptance $A(Q^2, W, \cos \theta_K^{c.m.}, \cos \theta_p^{RF})$ to weight each event. Given the small acceptance of CLAS12 for the 
$e'K^+p$ final state, the weighting for very small acceptance events can potentially distort the $MM(e'K^+)$ spectrum, which systematically 
affects the extracted weighted ({\it i.e.} acceptance-corrected) yields. To reduce the effect of these distortions, a minimum acceptance cut-off 
of 1\% of the average acceptance was applied.

Comparisons of the 2D and 4D acceptance models and analysis approaches are provided in Section~\ref{systematics}. The good agreement of the distinctly 
different approaches provides strong evidence for the reliability of the results and the integrity of the analysis chain.

\section{Yield Extraction}
\label{yields}

\begin{figure}[htbp]
\centering
\includegraphics[width=0.8\columnwidth]{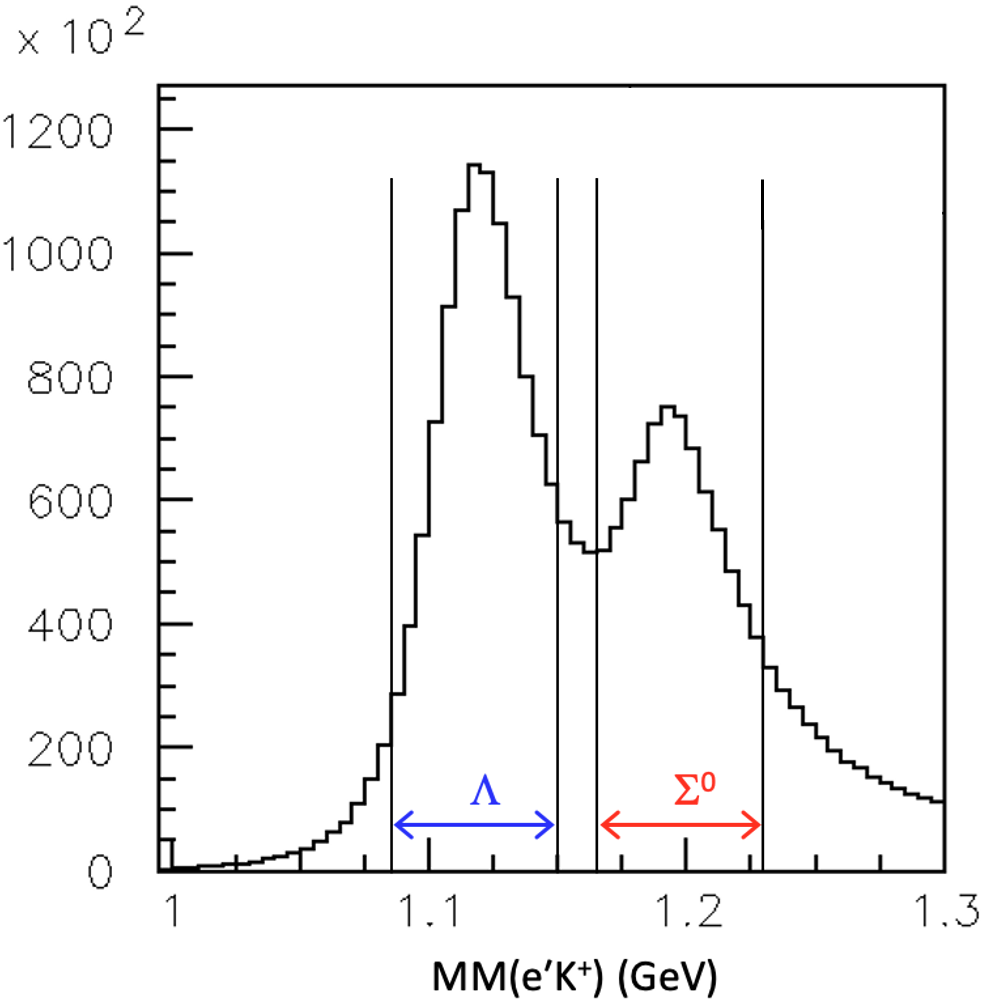}
\vspace{-2mm}
\caption{The $e'K^+$ missing mass distribution after all particle identification and exclusivity cuts for the 6.535~GeV dataset summed over all 
kinematics. The vertical lines about the $\Lambda$ and $\Sigma^0$ hyperon peaks identify the nominal analysis windows used to select the event 
samples.}
\label{mm-ranges}
\end{figure}

The hyperon recoil polarization was determined by analyzing data binned in the relevant kinematic variables $Q^2$, $W$, $\cos \theta_K^{c.m.}$, 
and $\cos \theta_p^{RF}$. For the $\Lambda$ and $\Sigma^0$ analyses, the different final states were selected in the $e'K^+$ missing mass 
distributions in mass regions about the individual hyperon peaks. The nominal $\Lambda$ mass region was chosen in the range from 1.090-1.150~GeV 
and the nominal $\Sigma^0$ mass region was chosen in the range from 1.165-1.230~GeV (see Fig.~\ref{mm-ranges}). The exact choices are somewhat 
arbitrary but were selected to maximize the event yields for the hyperons of interest while minimizing the contamination of the contributing 
backgrounds. See Section~\ref{systematics} for details on the systematic uncertainty regarding the chosen hyperon mass regions.

%%%%%%%%%%%%%%%%%%%%%%%%%%%%%%%%%%%%%%%%%%%%%%%%%%%%%%%%%%%%%%%%%%%%%%%%%%%%%%%%%%%%%%%%%%%%%%%%%%%%%%%%%%%%%%%%%%%%%%%%%%%%%%%%%%
%%%%%%%%%%%%%%%%%%%%%%%%%%%%%%%%%%%%%%%%%%%%%%%%%%%%%%%%%%%%%%%%%%%%%%%%%%%%%%%%%%%%%%%%%%%%%%%%%%%%%%%%%%%%%%%%%%%%%%%%%%%%%%%%%%

As shown in Fig.~\ref{mm-ranges} the hyperon signals in each mass region are not pure. Underlying both the $\Lambda$ and $\Sigma^0$ peaks is a 
background arising from multi-pion events dominated by the exclusive reaction channel $ep \to e'p\pi^+ \pi^-$, where the $\pi^+$ is misidentified 
by CLAS12 as a $K^+$ due to the finite timing resolution of the CLAS12 time-of-flight measurements. At momenta above $\sim$2.5~GeV in the FD and 
$\sim$0.8~GeV in the CD, the misidentification of $\pi^+$ tracks as $K^+$ allows the multi-pion topology to pollute the $K^+Y$ sample. In the 
$\Lambda$ mass region, the tail of the resolution-smeared $\Sigma^0$ peak contaminates the $\Lambda$ events. Within the $\Sigma^0$ mass region, there 
is a sizable contamination from $\Lambda$ radiative tail events. The cross contamination of the hyperons into the neighboring mass regions must be 
accounted for as the hyperons are polarized.

The approach to determine the three contributions to the $MM(e'K^+)$ spectrum relies on input from both MC and data. The hyperon contributions are 
accounted for by defined lineshape templates based on the realistic CLAS12 simulations (see Section~\ref{monte}). The kinematic range of the data used 
to parameterize the event generator output is well matched to the data included in this work and well reproduces the kinematic distributions vs. $Q^2$, 
$W$, and $\cos \theta_K^{c.m.}$. For this analysis, 200M events were generated for each of the $K^+\Lambda$ and $K^+\Sigma^0$ final states for each beam 
energy.

As the momentum resolution of the reconstructed MC for charged tracks is better than that for the data reconstruction, the MC $K^+Y$ template 
spectra were Gaussian smeared bin-by-bin in the mass spectra to minimize the $\chi^2$ in the template fits. The Gaussian smearing was optimized 
individually for each analysis bin in $Q^2$, $W$, $\cos \theta_K^{c.m.}$, and $\cos \theta_p^{RF}$. Within a given kinematic bin, the $\Lambda$ and 
$\Sigma^0$ Gaussian smearing parameters were constrained to be equivalent.

For the multi-pion background contribution, $ep \to e'p\pi^+X$ events from data were used with the $\pi^+$ re-assigned the $K^+$ mass. The 
same analysis code used for the $K^+Y$ events was used to sort the $MM(e'\pi^+)$ distributions in the same $Q^2$, $W$, $\cos \theta_K^{c.m.}$,
and $\cos \theta_p^{RF}$ kinematic bins.

\begin{figure*}[htbp]
\centering
\includegraphics[width=0.7\textwidth]{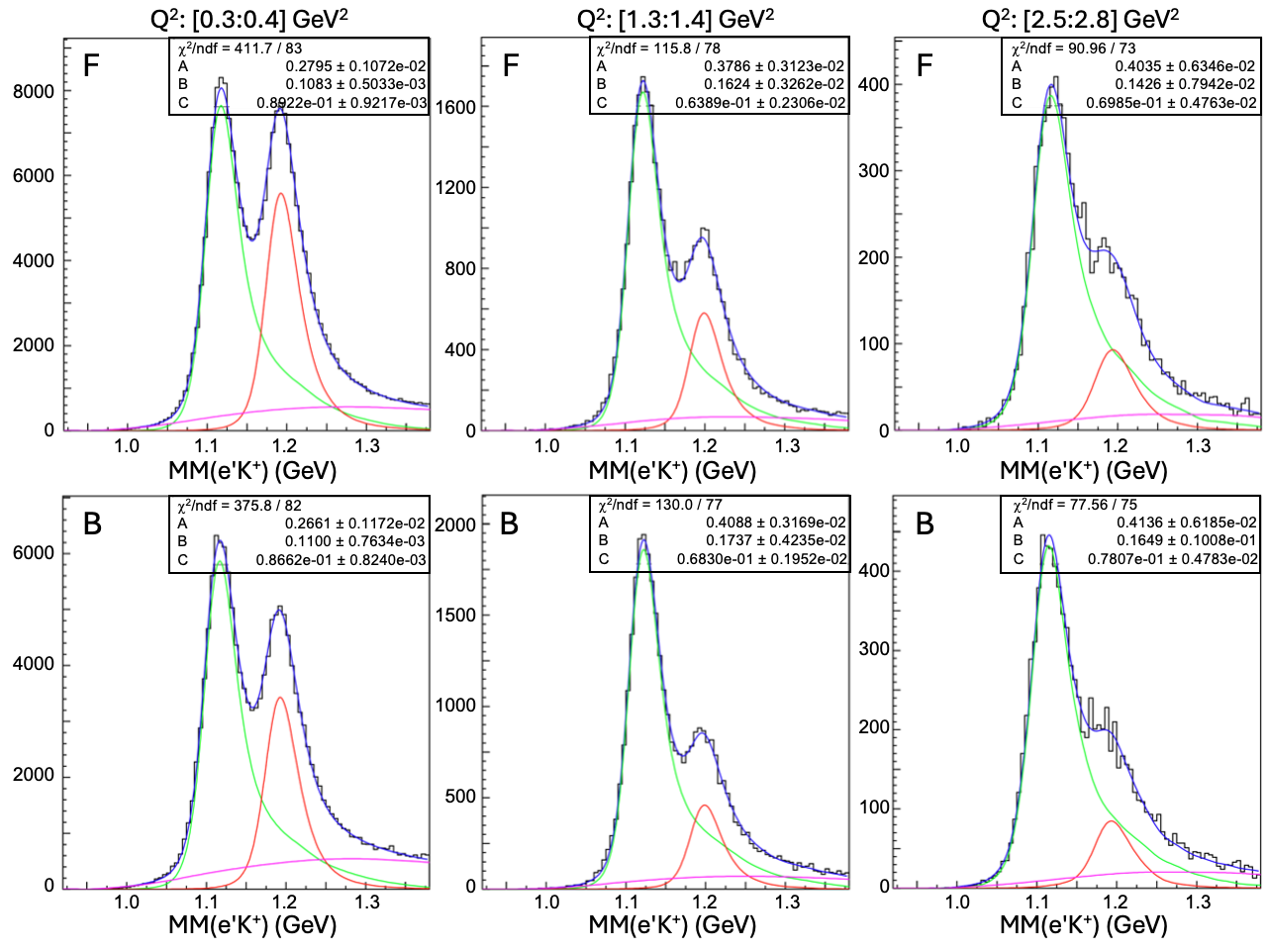}
\vspace{-4mm}
\caption{Representative $MM(e'K^+)$ spectrum fit results using hyperon templates derived from MC ($\Lambda$: green curve, $\Sigma^0$: red curve) and a 
background form based on multi-pion events where the $\pi^+$ has been misidentified as a $K^+$ based on measured data (magenta curve). The blue curve 
shows the full fit result. The fits shown here are from the 1D analysis for $F$ (top) and $B$ (bottom) bins at low, middle, and high $Q^2$ with respect 
to the $y'$ axis from the 6.535~GeV dataset.}
\label{mm-fits}
\end{figure*}

The $MM(e'K^+)$ spectrum in each analysis bin was then fit with a function of the form

\begin{equation}
MM = A \cdot \Lambda_{MC} + B \cdot \Sigma_{MC} + C \cdot BCK_{2\pi},
\end{equation}

\noindent
where $\Lambda_{MC}$ and $\Sigma_{MC}$ are the simulated hyperon distributions with weighting factors $A$ and $B$, respectively, and $BCK_{2\pi}$ 
is the template for the multi-pion background with a weighting factor $C$. Figure~\ref{mm-fits} shows representative spectrum fits to determine the 
hyperon yields. Systematic uncertainties for the fitting procedure were investigated varying the analysis range about the location of the $\Lambda$ 
and $\Sigma^0$ peaks - see Section~\ref{systematics} for details.

%%%%%%%%%%%%%%%%%%%%%%%%%%%%%%%%%%%%%%%%%%%%%%%%%%%%%%%%%%%%%%%%%%%%%%%%%%%%%%%%%%%%%%%%%%%%%%%%%%%%%%%%%%%%%%%%%%%%%%%%%%%%%%%%%%
%%%%%%%%%%%%%%%%%%%%%%%%%%%%%%%%%%%%%%%%%%%%%%%%%%%%%%%%%%%%%%%%%%%%%%%%%%%%%%%%%%%%%%%%%%%%%%%%%%%%%%%%%%%%%%%%%%%%%%%%%%%%%%%%%%

\begin{figure}[htbp]
\centering
\includegraphics[width=1.0\columnwidth]{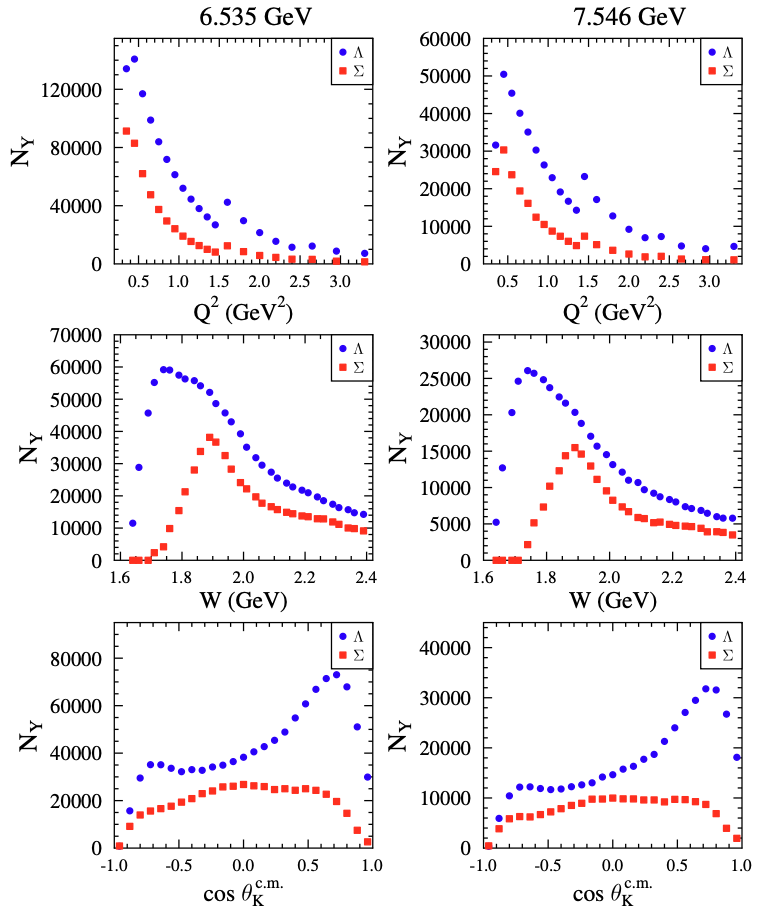}
\vspace{-6mm}
\caption{Hyperon yields from the 6.535~GeV and 7.546~GeV datasets vs. $Q^2$, $W$, and $\cos \theta_K^{c.m.}$ summed over the other two variables 
including all hadronic topologies. The data are limited to $Q^2$ in the range from 0.3 -- 3.5~GeV$^2$ for the 6.535~GeV dataset and in the range 
from 0.4 -- 4.5~GeV$^2$ for the 7.546~GeV dataset. In addition, all data are limited to $W$ from 1.6 -- 2.4~GeV. The blue (red) data points are 
for the $K^+\Lambda$ ($K^+\Sigma^0$) events in the $\Lambda$ ($\Sigma^0$) mass region. Note the discontinuity in the $Q^2$ plots in the top row
in the middle of the range is due to the bin-size change.}
\label{totals}
\end{figure}

Figure~\ref{totals} shows the determined $\Lambda$ yields in the $\Lambda$ mass region and the $\Sigma^0$ yields in the $\Sigma^0$ mass region for 
both the 6.535~GeV and 7.546~GeV datasets. These distributions are for the 1D polarization analysis (detailed in Section~\ref{binning}) sorting the 
polarization vs. $Q^2$, $W$, and $\cos \theta_K^{c.m.}$ integrated over the other two variables. The yields decrease rapidly with increasing $Q^2$ 
due to the roughly monopole fall-off of the kaon form factor. To compensate for this, the bin sizes were chosen to increase with $Q^2$. The yields vs. 
$W$ rise rapidly for the $K^+\Lambda$ and $K^+\Sigma^0$ channels within the first 100~MeV of their respective reaction thresholds peaking at 
$\sim$1.7~GeV for $K^+\Lambda$ and at $\sim$1.9~GeV for $K^+\Sigma^0$. The yields then gradually diminish with increasing $W$. The yields for 
$K^+\Lambda$ show a strong forward $\cos \theta_K^{c.m.}$ peaking due to the importance of $t$-channel kaon exchange contributions. As $K^+\Sigma^0$ 
has more $s$-channel contributions, it is peaked at central $\cos \theta_K^{c.m.}$ values. The very rapid fall-off just as $\cos \theta_K^{c.m.} \to 1$ 
is due to the forward acceptance hole of CLAS12 below $\theta \sim 5^\circ$. The recoil polarization analysis is based on separate $\Lambda$ and 
$\Sigma^0$ yield fits vs. $Q^2$, $W$, and $\cos \theta_K^{c.m.}$ for $\cos \theta_p^{RF} >0$ ($F$) and $\cos \theta_p^{RF} < 0$ ($B$) for the different 
coordinate axes $y'$ and $y$. 

%%%%%%%%%%%%%%%%%%%%%%%%%%%%%%%%%%%%%%%%%%%%%%%%%%%%%%%%%%%%%%%%%%%%%%%%%%%%%%%%%%%%%%%%%%%%%%%%%%%%%%%%%%%%%%%%%%%%%%%%%%%%%%%%%%
%%%%%%%%%%%%%%%%%%%%%%%%%%%%%%%%%%%%%%%%%%%%%%%%%%%%%%%%%%%%%%%%%%%%%%%%%%%%%%%%%%%%%%%%%%%%%%%%%%%%%%%%%%%%%%%%%%%%%%%%%%%%%%%%%%

The statistical uncertainties on the fitted yields $N_F$ and $N_B$ are given by the MINUIT~\cite{minuit} fit uncertainties and the uncertainties 
on the acceptance corrections $A_F$ and $A_B$ are based on binomial uncertainty propagation

\begin{equation}
A = \frac{N_{rec}}{N_{gen}}, \hskip 0.5cm \delta A = \frac{1}{N_{gen}} \sqrt{ N_{rec} (1 - A)},
\end{equation}

\noindent
where $N_{gen}$ is the number of generated MC events in a given kinematic bin and $N_{rec}$ is the corresponding number of reconstructed events in 
the bin. For all practical purposes, this contribution to the statistical uncertainty is negligible compared to the data uncertainty given the number 
of events employed for the simulation.

\section{Systematic Uncertainties}
\label{systematics}

In this section we examine the sources of systematic uncertainty that affect the extracted polarization observables. The procedure used to assign a 
systematic uncertainty to each source is to compare the polarization analysis results for all bins in $Q^2$, $W$, and $\cos \theta_K^{c.m.}$ with the
nominal analysis cuts/procedures and the modified cuts/procedures. A measure of the systematic uncertainty is the average difference between the 
extracted polarizations. For this analysis, our definition of the average polarization difference is defined by the root-mean square (RMS) width of 
the weighted polarization difference distribution given by

\begin{equation}
\label{sys-def}
\Delta {\cal P}^0 = \sqrt{ \frac{\sum\limits_i ({\cal P}^{0\,nom}_i - {\cal P}^{0\,alt}_i)^2/(\delta {\cal P}^0_i)^2}
  {\sum\limits_i 1/(\delta {\cal P}^0_i)^2}}.
\end{equation}

\noindent
Here the difference distribution sums over all data points, ${\cal P}^{0\,nom}_i$ (${\cal P}^{0\,alt}_i$) is the polarization of the $i^{th}$ point 
of the dataset with the nominal (altered) cuts/procedures in place and $\delta {\cal P}^0_i$ is the statistical uncertainty of the $i^{th}$ data 
point (using the nominal cuts/procedures) that is used as the weight factor.

The sources of systematic uncertainty in this analysis are separated into those that vary bin-by-bin and those that affect the average scale of the 
extracted ${\cal P}^0$ values. The contributions to the total systematic uncertainty belong to one of three general categories: i) polarization 
extraction, ii) acceptance function, and iii) background contributions. In the remainder of this section these different sources are reviewed.

\subsection{Polarization Extraction}
\label{pol_sys}

\noindent
\underline{Binning in $\cos \theta_p^{RF}$}: There is a systematic uncertainty contribution that arises due to the $\cos \theta_p^{RF}$ 
binning choice made during the data sorting. Nominally the data was sorted into two bins in $\cos \theta_p^{RF}$ referred to as forward 
and backward. A comparison of the polarization results with 6 $\cos \theta_p^{RF}$ bins was carried out through the full analysis chain 
with the 2D acceptance correction. With a smaller number of bins, the yields can be sensitive to the coarseness of the integrated acceptance 
corrections. Conversely, as the number of bins increases, the yield fits have larger uncertainties due to the decrease in the statistics in 
each bin. The RMS width of the weighted polarization difference distribution is 0.026, and this is assigned as a scale-type systematic 
uncertainty. The systematic for the 7.546~GeV data is consistent with that determined for the 6.535~GeV data although the binned yields in 
the smaller 7.546~GeV dataset have increased statistical uncertainties. 

\vskip 0.2cm
\noindent
\underline{Weak Decay Asymmetry Parameter}: A systematic uncertainty contribution arises due to the uncertainty in the weak decay asymmetry 
parameter $\alpha$. This uncertainty gives rise to a point-by-point type uncertainty on the extracted polarization (the same for both $\Lambda$ 
and $\Sigma^0$ hyperons) given by

\begin{equation}
  \delta {\cal P}^0_Y = \frac{\delta \alpha}{\alpha} \vert {\cal P}^0_Y \vert = \frac{0.009}{0.747}
  \vert {\cal P}^0_Y \vert = 0.012 \vert {\cal P}^0_Y \vert. 
\end{equation}

\subsection{Acceptance Function}
\label{acc_func}

There are several considerations that go into our final systematic uncertainty accounting associated with the acceptance correction 
employed in this analysis. These include the fiducial cuts on the DC and ECAL, the dimensionality and bin size of the acceptance table, 
and the matching of the EG model to the data.

\vskip 0.2cm
\noindent
\underline{Fiducial Cuts}: The nominal fiducial cuts for electrons and charged hadrons were designed to remove tracks at the azimuthal 
edges of the detectors. Studies of the extracted polarizations were made varying the DC fiducial cuts by $\Delta \phi = \pm 2^\circ$ on
each side of the individual detectors and varying the ECAL electron shower containment fiducial cuts by 5~cm on each side of the triangular
arrays in the 6 sectors. The average RMS width of the weighted difference distribution comparing nominal to loose and tight cuts is 0.004. 
This average is included as a point-by-point type uncertainty.

\vskip 0.2cm
\noindent
\underline{Acceptance Function ACC2D vs. ACC4D}: A comparison of the ${\cal P}^0$ results using the 2D acceptance model against the 4D 
acceptance model (see Section~\ref{monte}) resulted in RMS width of the weighted polarization difference for the $\Lambda$ analysis of 
0.016 and for the $\Sigma^0$ analysis of 0.055. Given the limited data statistics over the 4D cells ($Q^2$, $W$, $\cos \theta_K^{c.m.}$, 
$\cos \theta_p^{RF}$), coupled with the small CLAS12 acceptance for the $e'K^+p$ final state, the event-weighted $MM(e'K^+)$ distributions 
can show non-statistical fluctuations that affect the ${\cal P}^0$ extraction. Therefore the nominal analysis used the ACC2D acceptance
correction and the comparison of the ACC2D vs. ACC4D approaches was used to set a sysematic uncertainty for the acceptance correction.

\vskip 0.2cm
\noindent
\underline{EG Model}: To assign a systematic uncertainty for the EG model, the data-matched MC model was varied within the data uncertainties 
amounting to $\pm$5\%. The average RMS width of the weighted polarization difference is 0.009 for the $\Lambda$ and 0.032 for the $\Sigma^0$. 
As there is no clear kinematic dependence of the $\Delta {\cal P}^0$ variations, the RMS width values are assigned as a scale-type systematic 
uncertainty. Note that the comparisons of the ${\cal P}^0$ results with the MC acceptance based on the flat decay angle distributions in the 
hyperon decay frame to the modified model that matched the data gave RMS widths of the weighted polarization differences of 0.056 for the 
$\Lambda$ and 0.096 for the $\Sigma^0$ with the corresponding means of $\Delta {\cal P}^0$ of 0.118 for the $\Lambda$ and -0.067 for the 
$\Sigma^0$. These systematic shifts of the ${\cal P}^0$ results when applying a realistic physics model for the events shows the sensitivity 
of the results to the acceptance function.

\subsection{Background Contributions}
\label{bck_sys}

\noindent
\underline{$MM(e'K^+)$ Range}: Section~\ref{yields} shows the importance of the different background sources in the $\Lambda$ and $\Sigma^0$ 
analysis regions as a function of the kinematics. To check the level of stability of our results, the extent of the analysis regions about the 
hyperon peaks was increased by 5~MeV at each cut boundary to study the results employing looser $MM(e'K^+)$ cuts. For this study the 
complete analysis chain was repeated determining the corresponding CLAS12 acceptance function for each final state hyperon and extracting the 
yields within the modified $MM(e'K^+)$ range. The RMS width of the weighted difference distribution for the $\Lambda$ is 0.003 and for the 
$\Sigma^0$ is 0.013 for the 1D data sorts. These differences provide a good measure of the average systematic uncertainty for this contribution. 
It should be expected that the $\Sigma^0$ result is more sensitive to the definition of the analysis region due to the underlying $\Lambda$ 
contribution that gives rise to a larger systematic effect in the $\Sigma^0$ mass region. This uncertainty is assigned as a point-by-point 
type uncertainty.

\vskip 0.2cm
\noindent
\underline{Radiative Tail Modeling}: As discussed in Section~\ref{monte}, the generated MC $\Lambda$ and $\Sigma^0$ event samples were 
modified to include a modeling of the internal radiation contributions that are not included in either GEMC or the {\tt genKYandOnePion} 
EG. There is one free scaling parameter in the radiative model that was chosen to give the best overall $\chi^2$ in the $MM(e'K^+)$
fitting over all kinematic bins. To estimate the systematic uncertainty in the radiation model, the radiative tail scaling
parameter was varied over a very broad range by $\pm$50\% about the nominal value. The RMS width of the weighted polarization difference 
is 0.007 for the $\Lambda$ and 0.046 for the $\Sigma^0$. These uncertainties are assigned as point-to-point uncertainties using the average 
of the magnitudes. It should be expected that the uncertainties in the $\Sigma^0$ analysis are larger than for the 
$\Lambda$ analysis given that the $\Lambda$ radiative tail extends below the $\Sigma^0$ peak.

\vskip 0.2cm
\noindent
\underline{$MM^2(e'K^+p)$ Cut}: The $MM^2(e'K^+p)$ exclusivity cut is essential to cleanly select the $K^+\Lambda$ and $K^+\Sigma^0$ final states. 
By imposing this cut the backgrounds in the $MM(e'K^+)$ spectrum are significantly reduced. The nominal analysis cut limits on $MM^2(e'K^+p)$ were 
0-0.08~GeV$^2$. A study with altered cuts was carried out extending each cut boundary by 0.01~GeV$^2$. Varying this cut is an important study to 
ensure the modeling of the hyperon lineshapes with the modification of the radiative tail is well controlled. As well, it is important to show that 
the background contribution beneath the hyperons is understood and properly taken into account. The RMS width of the weighted polarization difference 
is 0.015 for the $\Lambda$ and 0.053 for the $\Sigma^0$. These values are assigned as the corresponding systematic uncertainties and are considered 
as a point-to-point systematic given the kinematic dependence present. It should be expected that the $\Lambda$ results are reasonably insensitive 
to the $MM^2(e'K^+p)$ cut variation as it mainly alters the $MM(e'K^+)$ spectrum above the $\Sigma^0$ peak location.

\vskip 0.2cm
\noindent
\underline{Detached Vertex Reconstruction Bias}: In the standard CLAS12 charged particle tracking algorithm in the FD and CD, the reconstruction
proceeds by swimming the track through the magnetic field back to the distance of closest approach to the beamline~\cite{recon-nim}. This procedure 
leads to a systematic misreconstruction of charged tracks that originate from a detached vertex. The effect of this bias can be studied with the MC. 
While $\Delta p_p = p_p^{true}-p_p^{rec}$ and $\Delta \theta_p = \theta_p^{true} - \theta_p^{rec}$ are centered at 0 with only resolution broadening, 
$\Delta \phi_p = \phi_p^{true} - \phi_p^{rec}$ shows a clear systematic misreconstruction with an asymmetric tail for $\Delta \phi_p > 0$. The bias is 
noticeably larger for the FD tracks compared to the CD tracks due to the significantly longer path length in the FD and the different magnetic field 
traversed (solenoid + toroidal).

The $\phi_p$ misreconstruction amounts to a few degrees in the CD but has a tail going out to $\Delta \phi_p \approx 20^\circ$ in the FD. Without 
detecting all final state particles, there is not enough information available to affect a proper correction. Instead the data and MC $\phi_p^{rec}$ 
distributions have been shifted to put the peak of $\Delta \phi_p$ from MC at zero for protons reconstructed in the FD and CD. This amounted to a 
2$^\circ$ shift for FD protons ($\phi_p^{rec,corr} = \phi_p^{rec} + 2^\circ$) and a 0.4$^\circ$ shift for CD protons 
($\phi_p^{rec,corr} = \phi_p^{rec} + 0.4^\circ$). 

To study the effect of this angle misreconstruction, the reconstructed $\phi_p$ angle was shifted by 3$^\circ$, corresponding to the average width
of the $\Delta \phi_p$ distribution. The full MC analysis chain was then carried out to extract ${\cal P}^0$ vs. $Q^2$, $W$, and $\cos \theta_K^{c.m.}$. 
The RMS width of the weighted polarization difference is 0.029 for the $\Lambda$ and 0.152 for the $\Sigma^0$. The $\Delta {\cal P}^0$ centroids are 
shifted to -0.033 for the $\Lambda$ and 0.096 for the $\Sigma^0$ and it is these values that are considered more indicative of the systematic uncertainty 
associated with this reconstruction bias. The magnitude of this offset was assigned as a scale-type uncertainty for the detached vertex charged track 
$\phi_p$ reconstruction bias as no apparent kinematic dependence was seen in $\Delta \phi_p$.

\vskip 0.2cm
\noindent
\underline{Beam Helicity Dependence}: The recoil hyperon polarization as detailed in Section~\ref{formalism} does not depend on the beam helicity. 
In this analysis the $\Lambda$ recoil polarization was studied separately for the $h = +1$ and $h = -1$ event samples. As the EG does not include a 
mechanism to model the hyperon polarization, the same CLAS12 acceptance corrections were employed for both the $h = +1$ and $h = -1$ events. The 
helicity-gated ${\cal P}^0$ values are consistent with each other to within 0.016, which is well within the statistical uncertainties of the measured 
polarizations, so no systematic uncertainty was assigned for this source.

\subsection{Final Systematic Uncertainty Accounting}

Our final systematic uncertainty accounting is included in Table~\ref{systab} listing all of the sources discussed above. The total systematic 
uncertainty, which represents the average over all kinematic bins, adds all of the individual contributions in quadrature assuming they are independent 
sources from each other. The systematic uncertainty for the background contributions is the dominant overall systematic source. Within this category 
the $\phi_p$ reconstruction bias associated with the detached vertex reconstruction is the biggest individual source. Note that with these assignments 
the systematic uncertainty is comparable to the statistical uncertainty across the full phase space. The average systematic uncertainty for the $\Lambda$ 
polarization is 0.049 and that for the $\Sigma^0$ is 0.138.

%%%%%%%%%%%%%%%%%%%%%%%%%%%%%%%%%%%%%%%%%%%%%%%%%%%%%%%%%%%%%%%%%%%%%%%%%%%%%%%%%%%%%%%%%%%%%%%%%%%%%%%%%%%%%%%%%%%%%%%%%%%%%%%%%%
\begin{table*}[htbp]
\begin{center}
\begin{tabular} {|c|c|c|c|} \hline
Category       & Contribution                                        & Systematic Uncertainty                & Category \\ \hline
Polarization   & Bin Size                     & 0.026                                 & Scale-type \\
Extraction     & Asymmetry Parameter          & 0.012$\cdot {\cal P}^0$               & Point-by-point \\ \hline
Acceptance     & Fiducial Cut Definition      & 0.004                                 & Point-by-point \\
Function       & Acceptance Table (2D vs. 4D) & 0.016 ($\Lambda$), 0.055 ($\Sigma^0$) & Scale-type \\
               & Model Dependence             & 0.009 ($\Lambda$), 0.032 ($\Sigma^0$) & Scale-type \\ \hline
Background     & $MM(e'K^+)$ Cut Limits       & 0.003 ($\Lambda$), 0.013 ($\Sigma^0$) & Point-by-point \\
Contributions  & Radiative Tail Modeling      & 0.007 ($\Lambda$), 0.046 ($\Sigma^0$) & Point-by-point \\
               & $MM^2(e'K^+p)$ Cut Limits    & 0.015 ($\Lambda$), 0.053 ($\Sigma^0$) & Point-by-point \\
               & $\phi_p$ Reconstruction Bias & 0.033 ($\Lambda$), 0.096 ($\Sigma^0$) & Scale-type \\ \hline
\multicolumn{4} {|r|} {{\bf $\langle$Total Systematic Uncertainty$\rangle$}: 0.049 ($\Lambda$), 0.138 ($\Sigma^0$)} \\ \hline
\end{tabular}
\caption{Summary table of the individual systematic uncertainty sources and the average total systematic uncertainty. Separate systematics were 
determined for the different hyperons. The right column indicates whether the systematic source has been categorized as a scale or a point-by-point 
type uncertainty.} 
\label{systab}
\end{center}
\end{table*}
%%%%%%%%%%%%%%%%%%%%%%%%%%%%%%%%%%%%%%%%%%%%%%%%%%%%%%%%%%%%%%%%%%%%%%%%%%%%%%%%%%%%%%%%%%%%%%%%%%%%%%%%%%%%%%%%%%%%%%%%%%%%%%%%%%

%%%%%%%%%%%%%%%%%%%%%%%%%%%%%%%%%%%%%%%%%%%%%%%%%%%%%%%%%%%%%%%%%%%%%%%%%%%%%%%%%%%%%%%%%%%%%%%%%%%%%%%%%%%%%%%%%%%%%%%%%%%%%%%%%%

\section{Results and Discussion}
\label{results}

\begin{figure*}[htbp]
\centering
\includegraphics[width=0.95\textwidth]{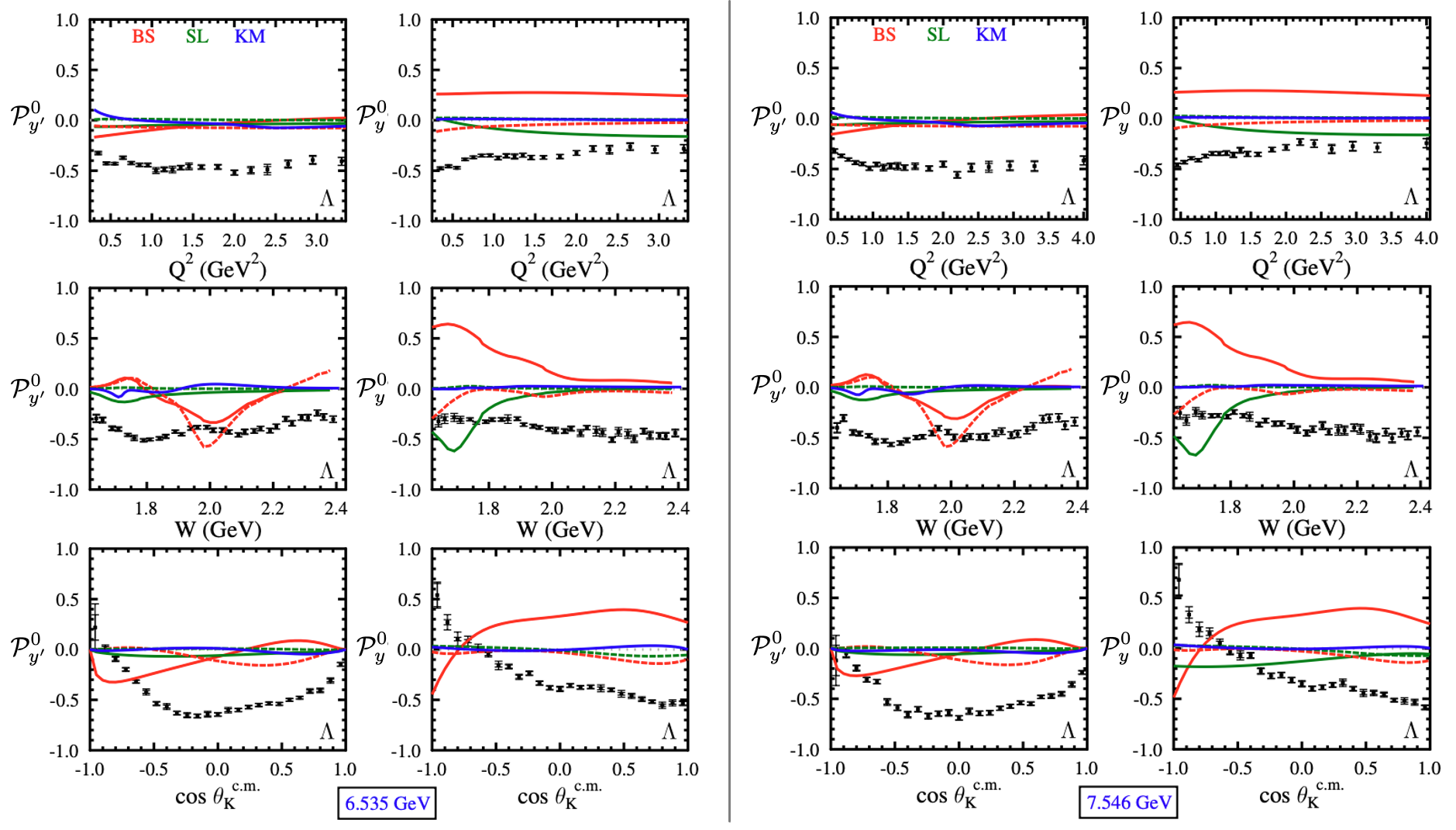}
\vspace{-3mm}
\caption{The $\Lambda$ recoil polarization components ${\cal P}^0$ with respect to the $y'$ and $y$ axes vs. $Q^2$ (top), $W$ (middle), and 
$\cos \theta_K^{c.m.}$ (bottom) for beam energies of 6.535~GeV (left) and 7.546~GeV (right). The data cover $Q^2$ from 0.3 (0.4) -- 3.5 (4.5)~GeV$^2$ 
for the 6.535~GeV (7.546~GeV) dataset and $W$ from 1.625 -- 2.4~GeV. In the text this is referred to as the 1D sort. The inner error bars are the 
statistical uncertainties and the outer error bars are the quadrature sum of the statistical and point-to-point systematic uncertainties. In addition 
there is a scale uncertainty of 0.046. The curves are calculations from the isobar models of Saclay-Lyon (SL)~\cite{saclay-lyon1,saclay-lyon2},
Kaon-MAID (KM)~\cite{kaon-maid1,kaon-maid2}, and the Czech group (BS)~\cite{skoupil18}. The dashed curves are from the models with the non-resonant 
terms turned off.}
\label{lpol1d}
\end{figure*}

\begin{figure*}[htbp]
\centering
\includegraphics[width=0.95\textwidth]{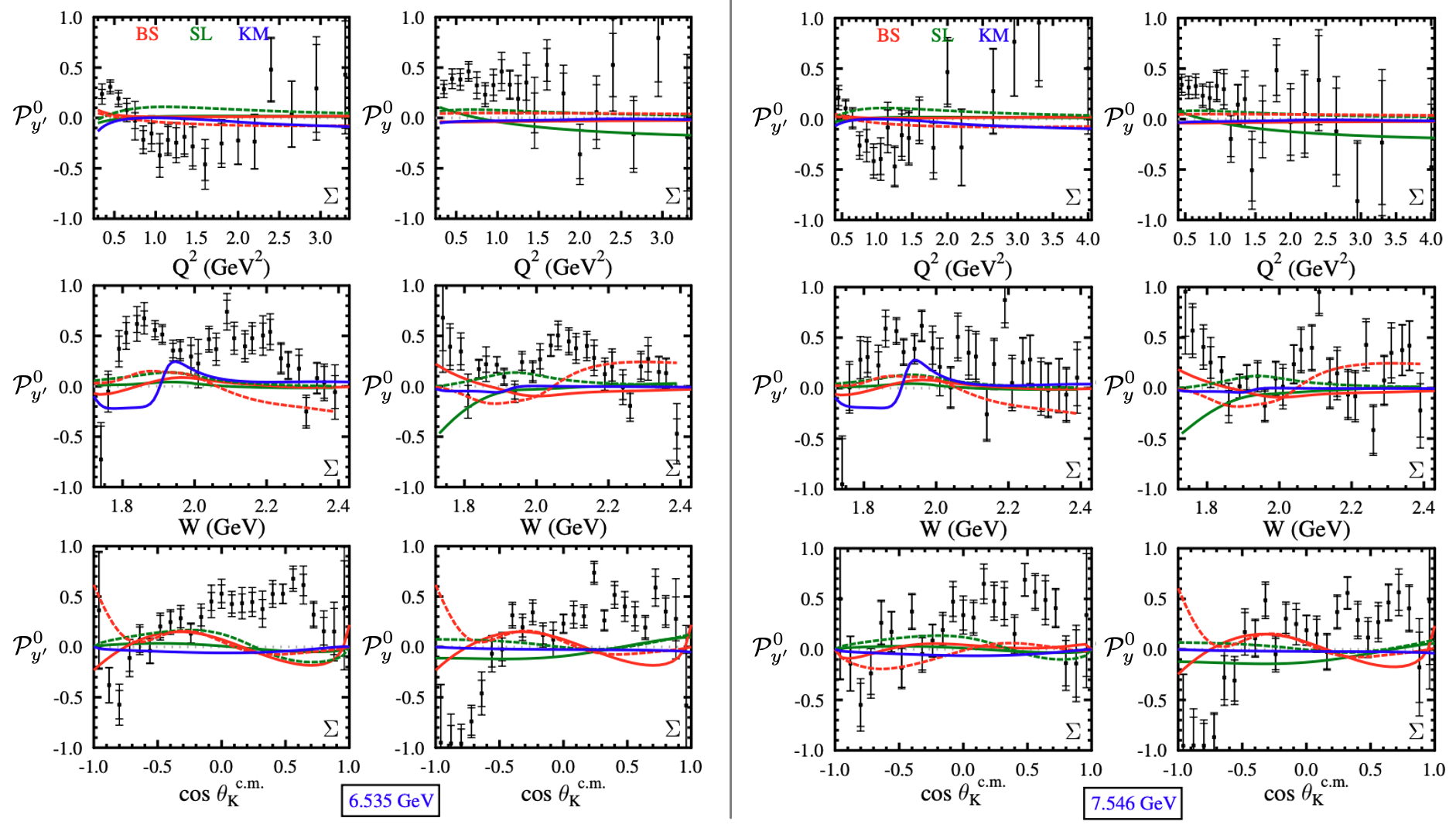}
\vspace{-3mm}
\caption{The $\Sigma^0$ recoil polarization components ${\cal P}^0$ with respect to the $y'$ and $y$ axes vs. $Q^2$ (top), $W$ (middle), and 
$\cos \theta_K^{c.m.}$ (bottom) for beam energies of 6.535~GeV (left) and 7.546~GeV (right). The data cover $Q^2$ from 0.3 (0.4) -- 3.5 (4.5)~GeV$^2$ 
for the 6.535~GeV (7.546~GeV) dataset and $W$ from 1.625 -- 2.4~GeV. In the text this is referred to as the 1D sort. The inner error bars are the 
statistical uncertainties and the outer error bars are the quadrature sum of the statistical and point-to-point systematic uncertainties. In addition 
there is a scale uncertainty of 0.118. The curves are calculations from the isobar models of Saclay-Lyon (SL)~\cite{saclay-lyon1,saclay-lyon2},
Kaon-MAID (KM) ~\cite{kaon-maid1,kaon-maid2}, and the Czech group (BS)~\cite{skoupil24}. The dashed curves are from the models with the non-resonant 
terms turned off.}
\label{spol1d}
\end{figure*}

\begin{figure*}[htbp]
\centering
\includegraphics[width=0.95\textwidth]{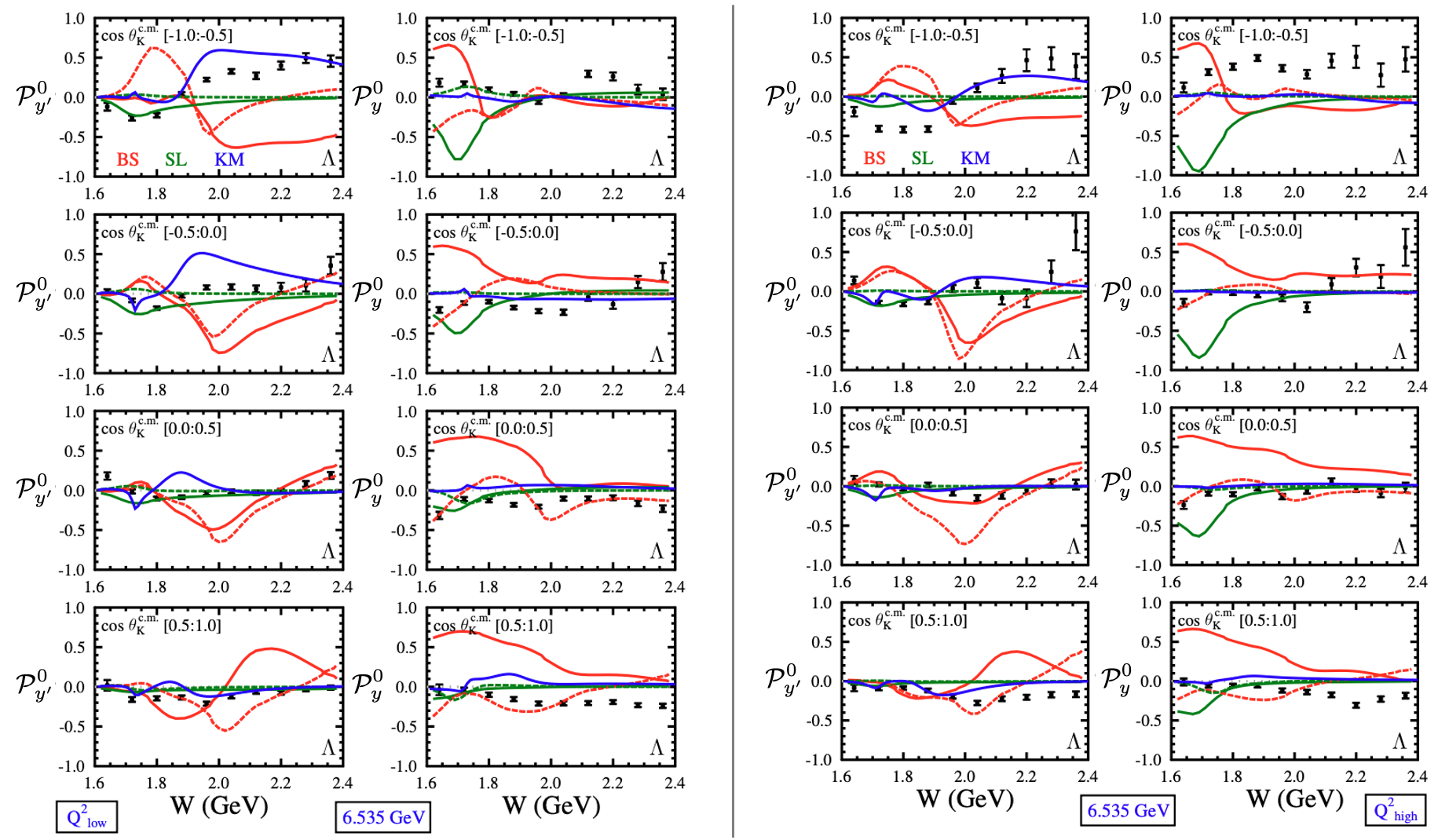}
\vspace{-3mm}
\caption{The $\Lambda$ recoil polarization components ${\cal P}^0$ for the 6.535~GeV dataset with respect to the $y'$ and $y$ axes vs. $W$ for four 
$\cos \theta_K^{c.m.}$ bins as labeled for $Q^2$ from 0.3 -- 0.9~GeV$^2$ (left) and $Q^2$ from 0.9 -- 3.5~GeV$^2$ (right). In the text this is referred 
to as the 3D sort. The inner error bars are the statistical uncertainties and the outer error bars are the quadrature sum of the statistical and 
point-to-point systematic uncertainties. In addition there is a scale uncertainty of 0.046. See the caption of Fig.~\ref{lpol1d} for a description of 
the model curves.}
\label{lpol3d}
\end{figure*}

\begin{figure*}[htbp]
\centering
\includegraphics[width=0.95\textwidth]{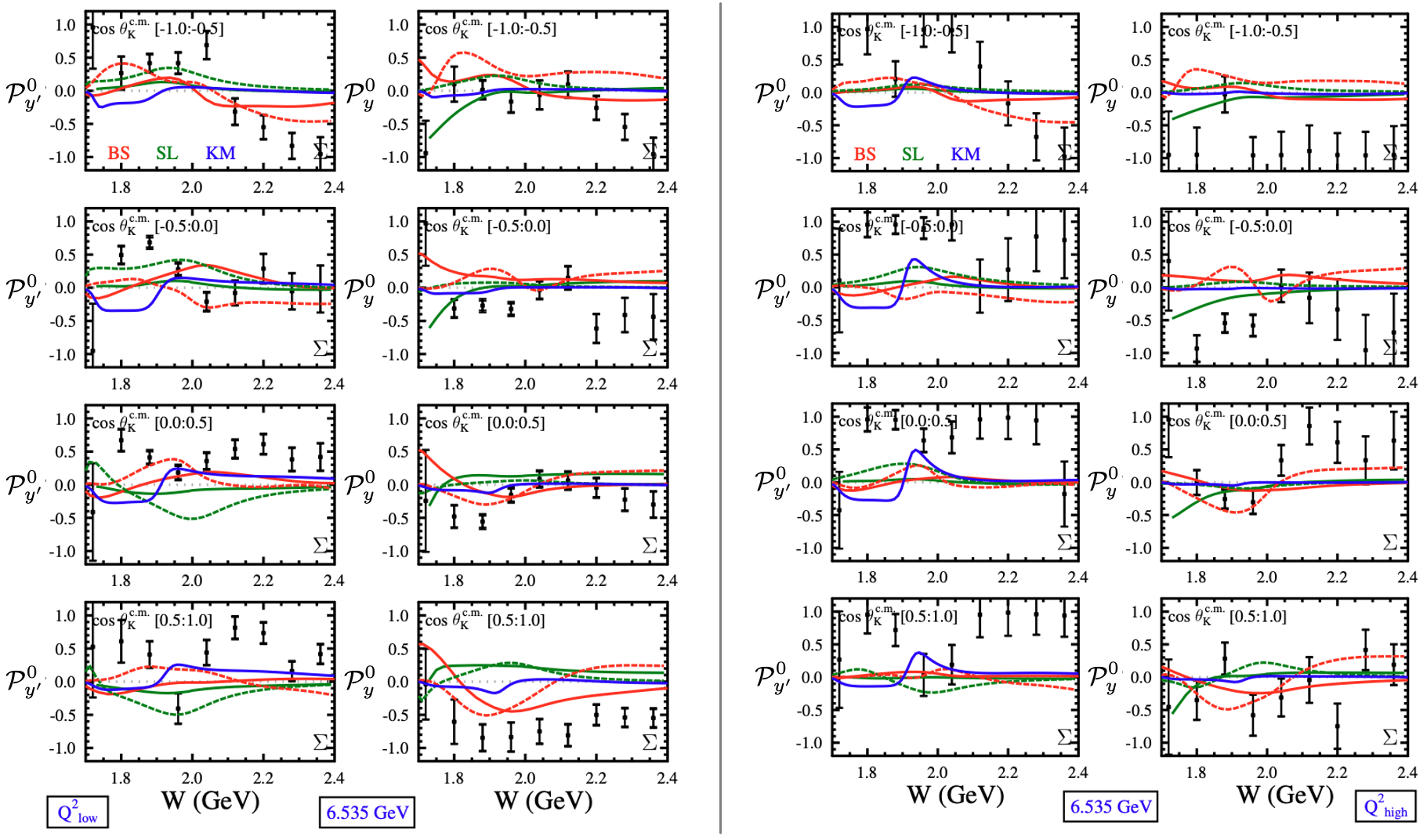}
\vspace{-3mm}
\caption{The $\Sigma^0$ recoil polarization components ${\cal P}^0$ for the 6.535~GeV dataset with respect to the $y'$ and $y$ axes vs. $W$ for four 
$\cos \theta_K^{c.m.}$ bins as labeled for $Q^2$ from 0.3 -- 0.9~GeV$^2$ (left) and $Q^2$ from 0.9 -- 3.5~GeV$^2$ (right). In the text this is referred 
to as the 3D sort. The inner error bars are the statistical uncertainties and the outer error bars are the quadrature sum of the statistical and 
point-to-point systematic uncertainties. In addition there is a scale uncertainty of 0.118. See the caption of Fig.~\ref{spol1d} for a description of 
the model curves.}
\label{spol3d}
\end{figure*}

Representative results for the recoil polarization ${\cal P}^0$ of the $\Lambda$ and $\Sigma^0$ hyperons in the exclusive $K^+\Lambda$ and $K^+\Sigma^0$ 
final states are shown for the 6.535~GeV and 7.546~GeV data in Figs.~\ref{lpol1d} and \ref{spol1d} for the 1D binning scenario and for the 6.535~GeV data 
in Figs.~\ref{lpol3d} and \ref{spol3d} for the 3D binning scenario. The error bars in these figures include statistical and point-to-point systematic 
uncertainties. The full set of the available data is included in the CLAS physics database~\cite{physicsdb}.

The ${\cal P}^0$ data for the $\Lambda$ hyperon vary rather smoothly vs. $Q^2$, $W$, and $\cos \theta_K^{c.m.}$. The most notable feature is in the
$W$ dependence of ${\cal P}^0_{y'}$ where statistically meaningful dips are seen at $W \approx 1.8$~GeV and $\approx 2.1$~GeV. The polarization has a 
slow and rather smooth variation vs. $Q^2$ with $\langle {\cal P}^0 \rangle \sim -0.5$. Both ${\cal P}^0_{y'}$ and ${\cal P}^0_y$ vary smoothly with 
$\cos \theta_K^{c.m.}$ in the range from $-0.5$ to 0.5, although their trends vs. $\cos \theta_K^{c.m.}$ are quite different. The data from the 6.535~GeV 
and 7.546~GeV datasets look quite similar, which is not unexpected given that the average $\epsilon$ values at a given $Q^2$ and $W$ are not so different 
for these two beam energies.

The data for ${\cal P}^0_\Sigma$ have larger statistical uncertainties compared to ${\cal P}^0_\Lambda$ due to the $\nu_\Sigma$ factor in the
polarization definition of Eq.(\ref{afb1}). The ${\cal P}^0_\Sigma$ components for the $y'$ and $y$ axes look quite similar in their dependence on
$\cos \theta_K^{c.m.}$, varying smoothly from $-0.5$ at back angles to 0.5 at forward angles. The dependence on $Q^2$ and $W$ displays a stronger
kinematic dependence compared to ${\cal P}^0_\Lambda$. Again the data are similar at 6.535~GeV and 7.546~GeV.

%%%%%%%%%%%%%%%%%%%%%%%%%%%%%%%%%%%%%%%%%%%%%%%%%%%%%%%%%%%%%%%%%%%%%%%%%%%%%%%%%%%%%%%%%%%%%%%%%%%%%%%%%%%%%%%%%%%%%%%%%%%%%%%%%%
%%%%%%%%%%%%%%%%%%%%%%%%%%%%%%%%%%%%%%%%%%%%%%%%%%%%%%%%%%%%%%%%%%%%%%%%%%%%%%%%%%%%%%%%%%%%%%%%%%%%%%%%%%%%%%%%%%%%%%%%%%%%%%%%%%

%%%%%%%%%%%%%%%%%%%%%%%%%%%%%%%%%%%%%%%%%%%%%%%%%%%%%%%%%%%%%%%%%%%%%%%%%%%%%%%%%%%%%%%%%%%%%%%%%%%%%%%%%%%%%%%%%%%%%%%%%%%%%%%%%%
%%%%%%%%%%%%%%%%%%%%%%%%%%%%%%%%%%%%%%%%%%%%%%%%%%%%%%%%%%%%%%%%%%%%%%%%%%%%%%%%%%%%%%%%%%%%%%%%%%%%%%%%%%%%%%%%%%%%%%%%%%%%%%%%%%

%%%%%%%%%%%%%%%%%%%%%%%%%%%%%%%%%%%%%%%%%%%%%%%%%%%%%%%%%%%%%%%%%%%%%%%%%%%%%%%%%%%%%%%%%%%%%%%%%%%%%%%%%%%%%%%%%%%%%%%%%%%%%%%%%%
%%%%%%%%%%%%%%%%%%%%%%%%%%%%%%%%%%%%%%%%%%%%%%%%%%%%%%%%%%%%%%%%%%%%%%%%%%%%%%%%%%%%%%%%%%%%%%%%%%%%%%%%%%%%%%%%%%%%%%%%%%%%%%%%%%

%%%%%%%%%%%%%%%%%%%%%%%%%%%%%%%%%%%%%%%%%%%%%%%%%%%%%%%%%%%%%%%%%%%%%%%%%%%%%%%%%%%%%%%%%%%%%%%%%%%%%%%%%%%%%%%%%%%%%%%%%%%%%%%%%%
%%%%%%%%%%%%%%%%%%%%%%%%%%%%%%%%%%%%%%%%%%%%%%%%%%%%%%%%%%%%%%%%%%%%%%%%%%%%%%%%%%%%%%%%%%%%%%%%%%%%%%%%%%%%%%%%%%%%%%%%%%%%%%%%%%

There are several different models shown in this work to compare against the polarization observables. The main features of the models are 
discussed here to set the stage for their comparisons to the data.

\vskip 0.2cm
\noindent
\underline{Saclay-Lyon} (green curves in Figs.~\ref{lpol1d} -- \ref{spol3d}) - The Saclay-Lyon (SL) model~\cite{saclay-lyon1,saclay-lyon2} is 
a tree-level isobar model that includes Born terms and $K^*(892)$ and $K_1(1270)$ exchanges in the $t$-channel. The model version used includes the 
$N(1440)1/2^+$, $N(1720)3/2^+$, and $N(1675)5/2^+$ $s$-channel resonances for both the $K^+\Lambda$ and $K^+\Sigma^0$ channels, along with the $u$-channel 
exchanges $\Lambda(1405)1/2^-$, $\Lambda(1670)1/2^-$, $\Lambda(1810)1/2^+$, and $\Sigma(1660)1/2^+$. In addition, the $K^+\Sigma^0$ channel includes 
exchanges of $\Delta(1900)1/2^+$, $\Delta(1232)3/2^+$, and $\Delta(1920)3/2^+$. The data used to constrain the parameters of the SL model were very limited 
given that it was developed before the release of any of the data produced from CLAS. The SL model calculations in Figs.~\ref{lpol1d} -- \ref{spol3d} 
include the full model and a version with the non-resonant terms switched off (no model refitting was done).

The SL model should not be expected to match the hyperon polarizations well given the lack of data available for constraints. However, the quality of its 
match to the data is no worse than later models developed based on fits to the photoproduction data from CLAS. Of course, without proper constraints from 
data at finite $Q^2$, there should be no expectation of good agreement from this archival model. The Saclay-Lyon calculations were provided by 
Ref.~\cite{czech-comm} and were integrated over the finite bins of this work.

\vskip 0.2cm
\noindent
\underline{Kaon-MAID} (blue curves in Figs.~\ref{lpol1d}-\ref{spol3d}) - The Kaon-MAID (KM) model is a tree-level isobar 
model~\cite{kaon-maid1,kaon-maid2} that includes Born terms, $K^*(892)$ and $K_1(1270)$ exchanges in the $t$-channel, and a limited set of spin 1/2, 3/2, and 5/2 
$s$-channel resonances. This core set includes the $N(1650)1/2^-$, $N(1710)1/2^+$, $N(1720)3/2^+$, and $N(1900)3/2^+$ for $K^+\Lambda$ and for 
$K^+\Sigma^0$ includes the same core set plus the $\Delta(1900)1/2^-$ and $\Delta(1910)1/2^+$. These states were chosen as they were reported to have 
non-zero decay widths into $K^+\Lambda$ and $K^+\Sigma^0$. The Born term vector meson exchange and the resonance couplings are based on fits to the available
$\gamma p \to K^+Y$ and $\pi^-p \to K^0\Lambda$ data. Kaon-MAID is not constrained by any 
$K^+Y$ electroproduction data. 

The Kaon-MAID model shows very sharp features in the $W$ dependence of ${\cal P}^0_\Lambda$ and ${\cal P}^0_\Sigma$ from the resonance terms for the
$y'$ axis at $\sim$1.7~GeV and 1.95~GeV, respectively, that are not seen in the data. However, the dependence of ${\cal P}^0$ vs.~$Q^2$ and 
$\cos \theta_K^{c.m.}$ varies smoothly. The model fails to describe the kinematic dependence of the data for either the $K^+\Lambda$ or the 
$K^+\Sigma^0$ channel. This model is archived online~\cite{kaon-maid} and the results included here were integrated over the finite bins of this work 
by its developer~\cite{mart-comm}.

\vskip 0.2cm
\noindent
\underline{Byd\v{z}ovsk\'{y}-Skoupil} (red curves in Figs.~\ref{lpol1d} -- \ref{spol3d}) - The Byd\v{z}ovsk\'{y}-Skoupil (BS) model~\cite{skoupil18,skoupil24} 
is another tree-level isobar model similar in design to the SL model. However, it represents a step forward in that it is based on fits to some of the 
available $\gamma p \to K^+\Lambda$ photoproduction data (differential cross sections, recoil polarization, beam spin asymmetry) and to some of the 
available $ep \to e'K^+\Lambda$ electroproduction data ($\sigma_U$, $\sigma_T$, $\sigma_L$, $\sigma_{LT'}$). However, it has not been constrained by any 
of the CLAS electroproduction data for the cross sections or structure functions detailed in Section~\ref{intro} that overwhelmingly dominate the available 
world data. The full set of 3- and 4-star PDG $N^*$ and $\Delta^*$ resonances of spins up to 5/2 and $W$ up to 2~GeV are included. In this work the
BS3 model was used for the $K^+\Lambda$ calculations.

Like other isobar models, the BS model includes Born terms and exchanges in the $t$- and $u$-channels to account for the non-resonant backgrounds. 
The work of Ref.~\cite{skoupil24} is an extension of the model from Ref.~\cite{skoupil18} to include the $K^+\Sigma^0$ final state. The $K^+\Sigma^0$ 
model of Ref.~\cite{skoupil24} is only for photoproduction. However, the electromagnetic vertices for the nucleon resonances and the Born terms include 
a dependence on $Q^2$. The model has been extended to electroproduction by including electromagnetic form factors to regularize the $Q^2$ behavior for 
large $Q^2$~\cite{czech-comm}. In this work the model-B version was used for the $K^+\Sigma^0$ calculations. The BS model calculations are shown in 
Figs.~\ref{lpol1d} -- \ref{spol3d} for the full model and with the non-resonant terms switched off (no model refitting done) to show the effect on the 
computed recoil polarizations. In this comparison, as for the SL model, the effects are quite striking given the differences in the curves.

The BS model, like the older SL isobar model, does not provide any better description of the data even though it includes significantly more
constraints based on fits to some of the available $KY$ photo- and electroproduction data. However, given that the response functions relevant 
for the recoil polarization in the BS model have not been constrained by any data, perhaps this is not so surprising. The comparisons of the model 
predictions to the data show that the model parameters for the form factors and coupling constants could be improved if the model were to include these 
new data as part of its constraints. All BS model calculations were provided by Ref.~\cite{czech-comm} and were integrated over the finite bins of 
this work.

\vskip 0.2cm
\noindent
\underline{JBW} - The J\"ulich-Bonn-Washington (JBW) coupled-channel model was initially developed to study $\pi$ and $\eta$ production~\cite{jbw-model}. 
It was recently extended to include most of the available CLAS $K^+\Lambda$ cross section data for photoproduction~\cite{jbw-kl1} and electroproduction
\cite{jbw-kl2} mentioned in Section~\ref{intro}. In this approach two-body unitary and analyticity are respected. Presently, the JBW model for $K^+\Lambda$ 
is limited to $W_{max} = 1.8$~GeV and $Q^2_{max} = 8$~GeV$^2$. The framework has also recently been extended to $K^+\Sigma^0$ photoproduction~\cite{jbw-ks} 
and a further extension to include the $K^+\Sigma^0$ electroproduction data is expected soon. To date no $K^+\Lambda$ polarization data at finite $Q^2$ are 
included in the fits to constrain the model. An initial comparison of the JBW model predictions to the CLAS12 $K^+\Lambda$ beam-recoil polarization transfer 
data provided in Ref.~\cite{carman-tpol} (see Ref.~\cite{jbw-kl2} Section 4) indicates that the hyperon polarization data in electroproduction data will be 
essential to constrain the contributions from the longitudinal multipoles.

\begin{figure}[htbp]
\centering
\includegraphics[width=1.0\columnwidth]{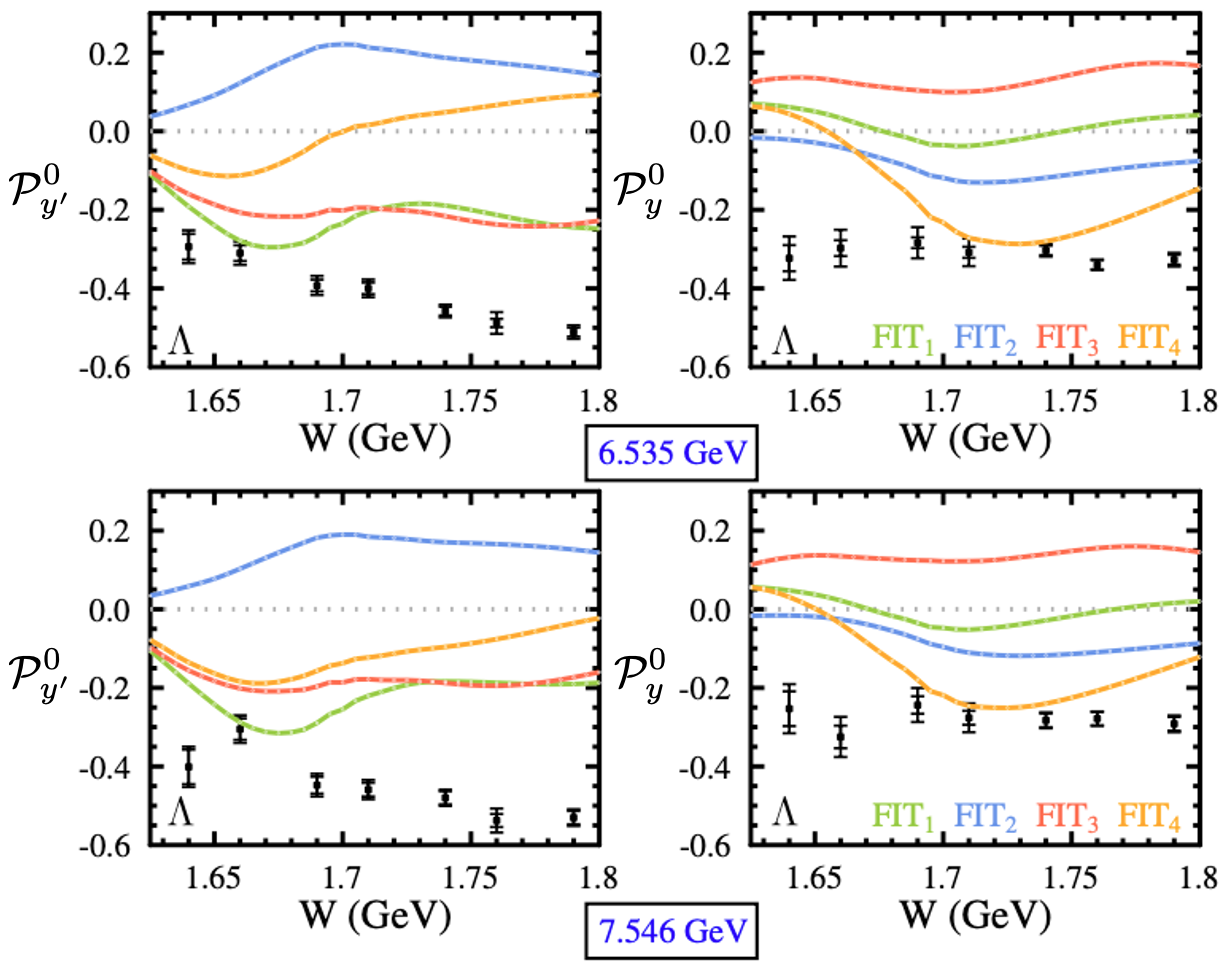}
\vspace{-7mm}
\caption{The $\Lambda$ recoil polarization components ${\cal P}^0$ with respect to the $y'$ and $y$ axes vs. $W$ for beam energies of 6.535~GeV (top) 
and 7.546~GeV (bottom) zoomed into the region of $W$ from 1.6-1.8~GeV from the 1D sort (as shown in Fig.~\ref{lpol1d}). The curves are calculations 
from the JBW coupled-channel model~\cite{jbw-kl2}. The four curves labeled $FIT_1 \to FIT_4$ are detailed in the text.}
\label{lpol1d-jbw}
\end{figure}

\begin{figure}[htbp]
\centering
\includegraphics[width=1.0\columnwidth]{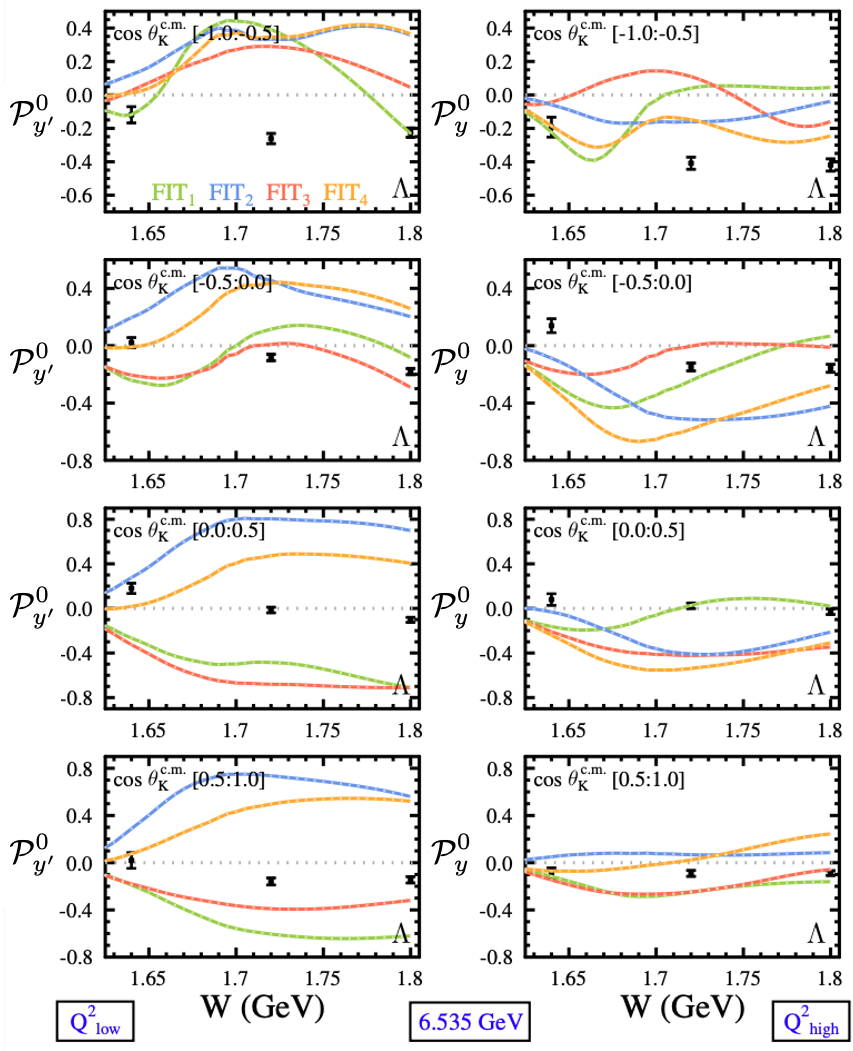}
\vspace{-7mm}
\caption{The $\Lambda$ recoil polarization components ${\cal P}^0$ for the 6.535~GeV dataset with respect to the $y'$ axis vs. $W$ zoomed into the region 
of $W$ from 1.6-1.8~GeV for four $\cos \theta_K^{c.m.}$ bins as labeled for $Q^2$ from 0.3 -- 0.9~GeV$^2$ (left) and $Q^2$ from 0.9 -- 3.5~GeV$^2$ (right) 
from the 3D data sort (as shown in Fig.~\ref{lpol3d}). The curves are calculations from the JBW coupled-channel model~\cite{jbw-kl2}. The four curves labeled 
$FIT_1 \to FIT_4$ are detailed in the text.}
\label{lpol3d-jbw}
\end{figure}

Figures~\ref{lpol1d-jbw} and \ref{lpol3d-jbw} show the ${\cal P}^0_\Lambda$ data compared to the predictions of the JBW model based on the available
data constraints. The calculations were provided by Ref.~\cite{mai-comm} and were integrated over the finite bins of this work. Figure~\ref{lpol1d-jbw}
shows the 1D data sort and Fig.~\ref{lpol3d-jbw} shows the 3D data sort vs. $W$ zoomed into the region from 1.6--1.8~GeV. The four JBW curves labeled
$FIT_1 \to FIT_4$ are detailed in Ref.~\cite{jbw-kl2}. $FIT_1/FIT_2$ and $FIT_3/FIT_4$ represent solutions from fits that converged to two different
local $\chi^2$ minima. $FIT_2$ and $FIT_4$ are refits of the $FIT_1$ and $FIT_3$ solutions, respectively, using a weighted $\chi^2$ definition to
account for the disparity of the much larger $\pi N$ dataset compared to the $K^+\Lambda$ dataset. 

The four JBW solutions do not compare favorably with the ${\cal P}^0$ data, with the data typically having a larger magnitude than the JBW predictions.
It is also interesting that the data shown in Fig.~\ref{lpol3d-jbw} are relatively independent of $Q^2$ while the JBW solutions have a very strong
$Q^2$ dependence.

%%%%%%%%%%%%%%%%%%%%%%%%%%%%%%%%%%%%%%%%%%%%%%%%%%%%%%%%%%%%%%%%%%%%%%%%%%%%%%%%%%%%%%%%%%%%%%%%%%%%%%%%%%%%%%%%%%%%%%%%%%%%%%%%%%
%%%%%%%%%%%%%%%%%%%%%%%%%%%%%%%%%%%%%%%%%%%%%%%%%%%%%%%%%%%%%%%%%%%%%%%%%%%%%%%%%%%%%%%%%%%%%%%%%%%%%%%%%%%%%%%%%%%%%%%%%%%%%%%%%%

%%%%%%%%%%%%%%%%%%%%%%%%%%%%%%%%%%%%%%%%%%%%%%%%%%%%%%%%%%%%%%%%%%%%%%%%%%%%%%%%%%%%%%%%%%%%%%%%%%%%%%%%%%%%%%%%%%%%%%%%%%%%%%%%%%
%%%%%%%%%%%%%%%%%%%%%%%%%%%%%%%%%%%%%%%%%%%%%%%%%%%%%%%%%%%%%%%%%%%%%%%%%%%%%%%%%%%%%%%%%%%%%%%%%%%%%%%%%%%%%%%%%%%%%%%%%%%%%%%%%%

\vskip 0.2cm

None of the models included here are able to reproduce the kinematic dependence seen in the data with their current parameters. These new electroproduction 
data from CLAS12, in addition to the full set of existing $K^+Y$ electroproduction cross section and polarization observables from CLAS detailed in
Section~\ref{intro}, can be expected to provide improved constraints. When the remainder of the data from the RG-K experiment is available in the near 
future, amounting to roughly a factor of ten increase from what is included here, much improved statistical precision with reduced bin sizes in $Q^2$, 
$W$, and $\cos \theta_K^{c.m.}$ will be possible. This will be important for further development of the different models.

\section{Summary and Conclusions}
\label{conclusions}

In this paper the hyperon recoil polarization for the exclusive electroproduction of the $K^+\Lambda$ and $K^+\Sigma^0$ final states from a proton 
target at beam energies of 6.535~GeV and 7.546~GeV are presented based on analysis of data from CLAS12 taken in Dec. 2018. The observables were 
measured in the range of $Q^2$ from 0.3-4.5~GeV$^2$ and $W$ from 1.6 to 2.4~GeV, while covering the full center-of-mass phase space of the final state 
$K^+$. The $\Lambda$ polarization measurements presented in this work extend the available data from the CLAS program. However, the data for the 
$\Sigma^0$ hyperon represent the first time this observable has been measured in electroproduction. This analysis of hyperon recoil polarization 
is a companion analysis to the beam-recoil hyperon transferred polarization analysis of Ref.~\cite{carman-tpol} based on analysis of this same dataset.

These new CLAS12 data have been compared to predictions from several available single-channel isobar models that have varying sensitivities to the 
$s$-channel resonance contributions. These different models do not compare favorably with the hyperon polarization data based on their current set of 
resonant/non-resonant diagrams and parameterizations. As well, the data have been compared to the predictions from an advanced coupled-channel model
based on fits to the available $K^+\Lambda$ photo- and electroproduction data in addition to the $\pi N$ and $\eta N$ data. All of these confrontations 
of the data to the models show that these new data from CLAS12 will allow for improved constraints on the models. It is also important to consider that 
reaction models whose development is based only on the fits to the available $\gamma p \to K^+Y$ photoproduction data cannot be expected to reproduce 
the electroproduction data. A proper reaction model will necessarily require a simultaneous fit to both $K^+Y$ photo- and electroproduction data over 
the broad kinematic range of the available data. Analyses of the CLAS12 data within a broad $Q^2$ range will allow us to establish the additional 
mechanisms contributing to $KY$ electroproduction that cannot be seen in photoproduction. These new mechanisms can be related either with the 
longitudinal electroproduction amplitudes or emerge gradually as $Q^2$ increases. Accounting for all mechanisms seen in the experimental data is 
critical for the extraction of the $\gamma_vpN^*$ electrocouplings.

These initial investigations into analysis of $KY$ data with CLAS12 will be greatly expanded in the near future with a new RG-K dataset collected in 
the first part of 2024 that is ten times larger than the 2018 dataset. The quality of the new data will allow for multi-dimensional extractions of
differential cross sections, separated structure functions, and polarization observables for the $K^+\Lambda$ and $K^+\Sigma^0$ final states. 
Ultimately, investigations of the exclusive production of $K^*Y$ and $KY^*$ final states will follow.

\begin{acknowledgments}

We thank Petr Byd\v{z}ovsk\'{y}, Jackson Hergenrather, Terry Mart, and Dalibor Skoupil for their efforts in preparing the model calculations for 
this paper. D.S.C. thanks Brian Raue for several useful discussions throughout the data analysis work. 

\vskip 0.2cm

We acknowledge the outstanding efforts of the staff 
of the Accelerator and the Physics Divisions at Jefferson Lab in making this experiment possible. This work was supported in part by the U.S.
Department of Energy, the National Science Foundation (NSF), and
the Italian Istituto Nazionale di Fisica Nucleare (INFN), 
the French Centre National de la Recherche Scientifique (CNRS), 
the French Commissariat pour l'Energie Atomique, 
the UK Science and Technology Facilities Council (STFC).
the National Research Foundation (NRF) of Korea, 
%the HelmholtzForschungsakademie Hessen f{\"u}r FAIR (HFHF), 
%the Chilean Agencia Nacional de Investigacion y Desarollo ANID PIA/APOYO AFB180002, 
%and 
the Skobeltsyn Nuclear Physics Institute and Physics Department at the Lomonosov Moscow State University. 
This work was supported in part by the the U.S. Department of Energy, Office of Science, Office of Nuclear Physics under contract DE-AC05-06OR23177.

\vskip 0.2cm
\noindent
$^\dag$ Corresponding author: carman@jlab.org

\end{acknowledgments}

\end{document}